\documentclass[aps,pre,twocolumn,superscriptaddress,longbibliography ,floatfix]{revtex4-2}
\usepackage{graphicx,bm,amssymb,amsmath,dcolumn,hyperref}
\usepackage{multirow}
\usepackage{enumerate}
\usepackage{enumitem}
\usepackage{color}
\usepackage{subfigure}  
\usepackage{epstopdf}
\usepackage{bbm}
\usepackage{graphicx}
\usepackage{hyperref}
\usepackage{threeparttable}
\usepackage{amsthm}
\usepackage{mathtools}
\usepackage{color}
\usepackage{tikz}
\usepackage{float}
\usepackage{empheq}

\begin{document}
\newcommand{\upcite}[1]{\textsuperscript{\textsuperscript{\cite{#1}}}}
\newcommand{\be}{\begin{equation}}
\newcommand{\ee}{\end{equation}}
\newcommand{\half}{\frac{1}{2}}
\newcommand{\ith}{^{(i)}}
\newcommand{\im}{^{(i-1)}}
\newcommand{\gae}
{\,\hbox{\lower0.5ex\hbox{$\sim$}\llap{\raise0.5ex\hbox{$>$}}}\,}
\newcommand{\lae}
{\,\hbox{\lower0.5ex\hbox{$\sim$}\llap{\raise0.5ex\hbox{$<$}}}\,}

\definecolor{blue}{rgb}{0,0,1}
\definecolor{red}{rgb}{1,0,0}
\definecolor{green}{rgb}{0,1,0}
\newcommand{\blue}[1]{\textcolor{blue}{#1}}
\newcommand{\red}[1]{\textcolor{red}{#1}}
\newcommand{\green}[1]{\textcolor{green}{#1}}
\newcommand{\orange}[1]{\textcolor{orange}{#1}}
\newcommand{\yd}[1]{\textcolor{blue}{#1}}

\newcommand{\scrA}{{\mathcal A}}
\newcommand{\scrB}{{\mathcal B}}
\newcommand{\scrE}{{\mathcal E}} 
\newcommand{\scrF}{{\mathcal F}} 
\newcommand{\scrL}{{\mathcal L}}
\newcommand{\scrM}{{\mathcal M}} 
\newcommand{\scrN}{{\mathcal N}}
\newcommand{\scrS}{{\mathcal S}}
\newcommand{\scrs}{{\mathcal s}}
\newcommand{\scrP}{{\mathcal P}}
\newcommand{\scrO}{{\mathcal O}}
\newcommand{\scrR}{{\mathcal R}}
\newcommand{\scrC}{{\mathcal C}}
\newcommand{\scrV}{{\mathcal V}}
\newcommand{\scrD}{{\mathcal D}}
\newcommand{\scrG}{{\mathcal G}}
\newcommand{\scrZ}{{\mathcal Z}}
\newcommand{\scrW}{{\mathcal W}}
\newcommand{\Lm}{\rm L_{\rm m}}

\newcommand{\PP}{\mathbb{P}}
\newcommand{\ZZ}{\mathbb{Z}}
\newcommand{\EE}{\mathbb{E}}
\renewcommand{\d}{\mathrm{d}}
\newcommand{\dm}{d_{\rm min}}
\newcommand{\rhojunction}{\rho_{\rm j}}
\newcommand{\rhojunctionLim}{\rho_{{\rm j},0}}
\newcommand{\rhobranch}{\rho_{\rm b}}
\newcommand{\rhobranchLim}{\rho_{{\rm b},0}}
\newcommand{\rhononbridge}{\rho_{\rm n}}
\newcommand{\rhononbridgeLim}{\rho_{{\rm n},0}}
\newcommand{\percolationCluster}{C}
\newcommand{\leafFreeCluster}{C_{\rm \ell f}}
\newcommand{\bridgeFreeCluster}{C_{\rm bf}}
\newcommand{\df}{d_\textsc{f}}
\newcommand{\yt}{y_{\rm t}}
\newcommand{\yh}{y_{\rm h}}
\newcommand{\dB}{d_\textsc{B}}
\newcommand{\dfprime}{d'_{\rm f}}
\newcommand{\bfx}{{\bf x}}

\newcommand{\sC}{\mathcal{C}}

\title{Backbone and shortest-path exponents  of the  two-dimensional $Q$-state Potts model}
\date{\today}
\author{Sheng Fang}
\affiliation{MinJiang Collaborative Center for Theoretical Physics,
	College of Physics and Electronic Information Engineering, Minjiang University, Fuzhou 350108, China}
\affiliation{Department of Modern Physics, University of Science and Technology of China, Hefei, Anhui 230026, China}	

\author{Da Ke}
\affiliation{MinJiang Collaborative Center for Theoretical Physics,
	College of Physics and Electronic Information Engineering, Minjiang University, Fuzhou 350108, China}
\author{Wei Zhong}
\email{w.zhong@mju.edu.cn}
\affiliation{MinJiang Collaborative Center for Theoretical Physics,
	College of Physics and Electronic Information Engineering, Minjiang University, Fuzhou 350108, China}
\author{Youjin Deng}
\email{yjdeng@ustc.edu.cn}
\affiliation{MinJiang Collaborative Center for Theoretical Physics,
	College of Physics and Electronic Information Engineering, Minjiang University, Fuzhou 350108, China}
\affiliation{Department of Modern Physics, University of Science and Technology of China, Hefei, Anhui 230026, China}	
\affiliation{
Shanghai Research Center for Quantum Sciences, Shanghai 201315, China}

\begin{abstract}
We present a Monte Carlo study of the backbone and the shortest-path exponents of 
the two-dimensional $Q$-state Potts model in the Fortuin-Kasteleyn bond representation. 
We first use cluster algorithms to simulate the critical Potts model on the square lattice 
and obtain the backbone exponents $\dB = 1.732 \, 0(3)$ and $1.794(2)$ for $Q=2,3$, respectively. 
However, for large $Q$, the study suffers from serious critical slowing down and 
slowly converging finite-size corrections. 
To overcome these difficulties, we consider the O$(n)$ loop model on the honeycomb lattice 
in the densely packed phase, which is regarded to correspond to the critical Potts model with $Q=n^2$.
With a highly efficient cluster algorithm, we determine from domains enclosed by the loops 
$\dB =1.643\,39(5), 1.732\,27(8), 1.793\,8(3),  1.838\,4(5), 1.875\,3(6)$ 
for $Q = 1, 2, 3, 2 \! + \! \sqrt{3}, 4$, respectively, 
and $\dm = 1.094\,5(2), 1.067\,5(3), 1.047\,5(3), 1.032\,2(4)$ for $Q=2,3, 2+\sqrt{3}, 4$, respectively. 
Our estimates significantly improve over the existing results for both $\dB$ and $\dm$.
Finally, by studying finite-size corrections in backbone-related quantities, 
we conjecture an exact formula as a function of $n$ for the leading correction exponent.
\end{abstract}
\maketitle

\section{Introduction}
\label{Introduction}

The Potts model \cite{potts1952some,wu1982potts} is an extension of the celebrated Ising model and 
plays an important role in the theory of phase transition and  critical phenomena. 
It also has broad applications in various fields like condensed-matter physics~\cite{wu1982potts}.
Given a connected graph $G \equiv (V,E)$ with $V$ the vertex (site) set and $E$ the edge set, 
each vertex has a spin of an integer value $\sigma = 0,1,\cdots , Q-1$, 
and two spins at the ends of an edge are coupled as $-J \delta_{\sigma_i, \sigma_j}$, 
where $J$ represents the interaction strength and $\delta$ is the Kronecker delta function.  
Accordingly, the partition function can be written as 
\begin{equation}
\label{eq:Potts}
\scrZ_{\rm Potts} = \sum_{\{ \sigma \}} \prod_{\langle ij \rangle } e^{J \delta_{\sigma_i, \sigma_j}},
\end{equation}
where the inverse temperature $\beta$ has already been set to be $1$, the summation $\{ \sigma \}$ is over all possible spin configurations,
and $\langle ij \rangle$ is for all pairings of spins on the edges. 
For $Q=2$, the Potts model reduces to the Ising model with the Ising coupling strength $K=J/2$.

The Potts model can be reformulated in graphical representations, 
including the Fortuin-Kasteleyn (FK) bond~\cite{kasteleyn1969phase,Grimmett2006} 
and $Q$-flow (loop) representations~\cite{Essam1986the,wu1988potts}. 
Both the graphical representations can be obtained from the original spin representation 
by high-temperature expansion techniques,
and, recently, these two graphical models also have been directly mapped onto each other 
with the introduction of a loop-cluster joint model \cite{zhang2020loop}. 
\par 

The FK bond representation is also known as the random-cluster (RC) representation, 
in which each edge is either empty or occupied.
Each occupied bond has a statistical weight (relative to an empty one) as $v=\exp(J)-1$,
and each connected component (also called cluster) has a fugacity $Q$. 
The partition function of the RC model then reads as 
\begin{equation}
\scrZ_\textsc{rc} = \sum_{\scrA \subseteq G} v^{\scrN_b} Q^{\scrN_c} \; ,
\label{eq:RC}
\end{equation}
where the summation is over all possible subgraphs of $G$, and $\scrN_b$ and $\scrN_c$ 
represent the total numbers of occupied bonds and clusters, respectively. 
As an important consequence, the parameter $Q$ can now take any nonnegative real number. 
For ferromagnetic coupling $J \! > \! 0$, the bond weight $v$ is positive,
and each configuration in Eq.~(\ref{eq:RC}) has a probability interpretation. 
As special cases, 
the RC model reduces to the standard bond percolation in the $Q \! \rightarrow \! 1$ limit,
and the spanning tree or forest for $Q \! \rightarrow \! 0$ \cite{Grimmett2006}. 
The FK representation has an important role in conformal field theory \cite{francesco2012conformal}
and in stochastic Loewner evolution \cite{kager2004guide,cardy2005sle}, 
leading to much advanced theoretical progresses for the Potts model. 
Further, based on passing back and forth between the spin and FK representations, 
efficient cluster methods (i.e., the Swendsen–Wang (SW) \cite{swendsen1987nonuniversal} and
Wolff algorithms \cite{wolff1989collective}) are developed and widely used.

In addition to physical quantities from the original spin representation, 
the RC model has very rich geometric structures associated with FK random clusters,
and a variety of critical exponents are used to characterize these geometric behaviors, which were first introduced in percolation \cite{herrmann1984building,herrmann1984backbone,stauffer2018introduction}. 
The Euclidean diameter $\xi_1$ of the largest cluster acts as the correlation length
and diverges as $\xi_1 \! \sim \! t^{-\nu}$, with $t \! \equiv \! (v_c-v)/v_c$ and $v_c$ the critical point,
and $\nu$ is frequently called the correlation-length exponent. 
The mass $s$ of any critical cluster has a power-law dependence 
on its diameter $\xi_s $ as $s \sim \xi_s^{\df}$ for $s \gg 1$, with $\df$ the fractal dimension.
In two dimensions (2D), the hull and the external perimeter of clusters
can be further defined and have fractal dimensions $d_{\rm hull}$ and $d_\textsc{ep}$, respectively.  
For a pair of connected sites with distance $|\mathbf{x}| \gg 1 $, 
the graph distance $\mathcal{S}$, i.e., the minimum length of all connecting paths between them, 
algebraically diverges as $\mathcal{S} \sim |\mathbf{x}|^{\dm}$, with $\dm$ the shortest-path exponent. 
If a voltage difference is applied between these two sites, 
the total number $N_\textsc{b}$ of bonds carrying nonzero current
scales as $ N_\textsc{b} \sim |\mathbf{x}|^{\dB}$, where $\dB$ is called the backbone exponent. 
Further, the total number $N_{\rm red}$ of red bonds, which carry all the current, 
behaves as  $N_{\rm red} \sim |\mathbf{x}|^{d_{\rm red}}$, 
with $d_{\rm red}$ the red-bond exponent. 

In the past decades, the $Q$-state Potts model has been extensively studied~\cite{baxter2016exactly,wu2009exactly,nienhuis1984critical,nienhuis1987coulomb}.
In 2D, the Potts model exhibits a second-order phase transition for $Q \! \le \! 4$,
and the exact values of most critical exponents have been identified.
In the Coulomb gas theory \cite{nienhuis1984critical,nienhuis1987coulomb}, the Coulomb-gas coupling strength $g$ relates to $Q$ as 
\begin{equation}
\label{eq:CG-coupling}
Q = 4 \cos^2(\pi g/4) \; , \hspace{5mm} g \in [2,4],
\end{equation}
and the thermal and magnetic exponents of the leading and subleading renormalization fields are known as 
\begin{align}
    \label{eq:CGprediction1}
    & y_{\rm t1} = 3-6/g  \; , \qquad \qquad \; \: y_{\rm t2} = 4-16/g  \; ,   \\
    & y_{\rm h1} = \frac{(g+2)(g+6)}{8g} \,, \;\;\;\;\;  y_{\rm h2} = \frac{(g+10)(g-2)}{8g} \; ,   \nonumber
\end{align}
where the leading thermal exponent $y_{\rm t1}$ is the inverse of 
the correlation-length exponent $\nu$ as $y_{\rm t1}=1/\nu$.
The subleading field, with $y_{\rm t2} \! < \! 0$, governs the convergence of leading corrections;
logarithmic corrections arise as $Q$ approaches to 4, for which $y_{\rm t2} \! \rightarrow \! 0$.

The fractal dimension $\df$ of FK clusters is identical to the leading magnetic 
exponent as $\df = y_{\rm h1}$,
and the exact values of some other geometric exponents are \cite{coniglio1989fractal} 
\begin{align}
    \label{eq:CGpredictions}
    & d_\textsc{\rm hull} = 1+2/g \; ,  \hspace{20mm}   d_\textsc{ep} = 1+g/8 \; , \nonumber \\
    & d_{\rm red} \: =  (4-g)(3g + 4)/g \; . 
\end{align}
However, there is still a set of geometric critical exponents, including $\dB$ and $\dm$, whose exact values are unavailable. 
Monte Carlo simulations have been used to estimate them  \cite{deng2004backbone,deng2010some,xu2014geometric,hou2019geometric,elcci2016bridges,zhou2012shortest},
and Table~\ref{tab:estimate_result} lists some estimates of $\dB$ and $\dm$. \par 

In this work, we apply the SW and Wolff cluster methods 
to simulate the critical Potts model on the square lattice, 
expecting to obtain better results for $\dB$ and $\dm$ than those in Refs.~\cite{deng2004backbone,xu2014geometric,hou2019geometric,zhou2012shortest,deng2010some}. 
An improved algorithm is formulated to classify the occupied bonds into ``bridges" and ``nonbridges" \cite{xu2014geometric}.
From the fractal dimension of bridge-free clusters, we determine  $\dB =1.732\,0(3)$ for $Q=2$ 
and $1.794(2)$ for $Q=3$. 
The estimate of $\dB$ for $Q=2$ is consistent with Ref.~\cite{hou2019geometric}, 
and for $Q=3$ it nearly rules out the result in Ref.~\cite{deng2004backbone}. 
As $Q$ increases, however, 
the cluster methods suffer from severe critical slowing-down
and finite-size corrections become very strong,
preventing us to obtain high precision of $\dB$ for large $Q$. \par 

\begin{table}[t]
    \centering
    \begin{tabular}{|p{1.9cm}|p{2.5cm}|p{2.5cm}|}
    \hline 
        ($Q$,$g$) & $\dB$ &$\dm$  \\
    \hline 
        (1,8/3)& 1.643\,36(10)\cite{xu2014geometric} & 1.130\,77(2)\cite{zhou2012shortest}   \\
        present  & 1.643\,39(5)    &         \hspace{3mm} -          \\
        \hline 
        (2,3) & 1.732\,1(4)\cite{hou2019geometric} & 1.094\,0(2) \cite{hou2019geometric} \\
        present  & 1.732\,27(8)                & 1.094\,5(2)                      \\
        \hline 
        (3,10/3) & 1.789\,5(5)\cite{deng2004geometric}& 1.066\,2(30) \cite{deng2010some} \\
        present  &  1.793\,8(3)                & 1.067\,5(3)                       \\
        \hline 
        (2+$\sqrt{3}$,11/3) &\hspace{3mm} -           &  \hspace{3mm} -           \\
        present  &  1.838\,4(5)                & 1.047\,5(3)                   \\
        \hline 
        (4,4) & 1.873(4)\cite{deng2004geometric}  & \hspace{3mm} -   \\
        present  & 1.875\,3(6)                & 1.032\,2(4)         \\
        \hline 
    \end{tabular}
    \caption{Best estimates of $\dB$ and $\dm$ for the two-dimensional Potts model with $Q=1,2,3,2+\sqrt{3},4$ and $g$ is the Coulomb-gas coupling strength. These final quoted values and the error margins are estimated from a variety of fitting results by taking into account statistical and systematic errors.}
    \label{tab:estimate_result}
\end{table}

We then consider the O$(n)$ loop model \cite{domany1981duality} on the honeycomb lattice,
of which the configuration is a gas of nonintersecting loops.
Let $n$ be the fugacity of a loop and $x$ be the statistical weight 
for a loop segment (an occupied bond); the partition function reads as
\begin{align}
\label{eq:ON_loop_PF}
\scrZ_{\rm loop} = \sum_{\rm loops } x^{\scrL} n^{\scrN_\ell} \; ,
\end{align}
where $\scrN_\ell$ is the number of loops and $\scrL$ is the total length of all the loops. 
For $n=1$, the loop model reduces to the flow (loop) representation of 
the honeycomb-lattice Ising model, with the coupling strength as $\tanh (K)=x$ \cite{thompson2015mathematical}.

\begin{figure}[t]
    \centering
    \includegraphics[width=0.50\textwidth]{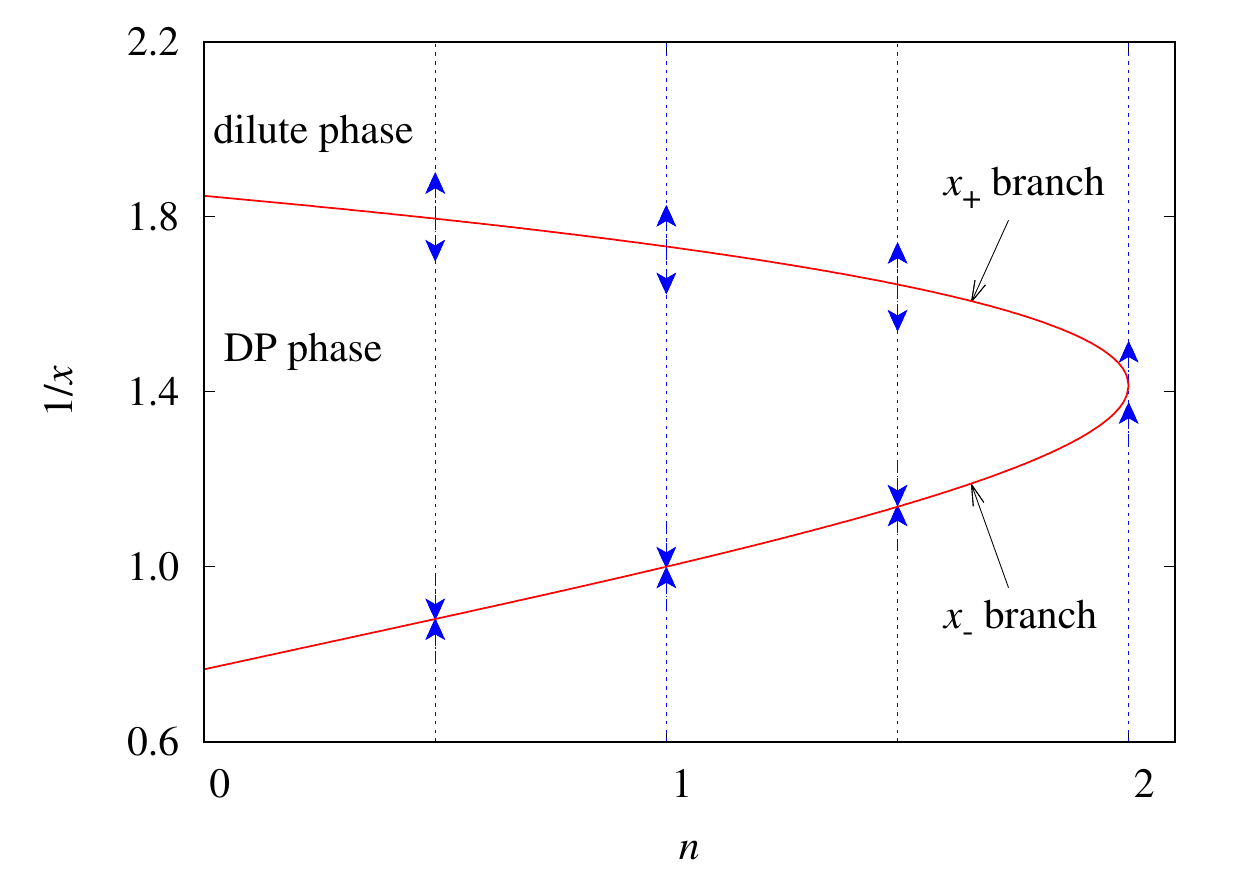}
    \caption{Phase diagram of the O($n$) loop model on the honeycomb lattice.
    The system is in dilute phase if bond weight $x \! < \! x_{+}$ 
    and in densely packed (DP) phase if $x \!>\! x_{+}$.
    Vertical blue arrows sketch the directions of renormalization flows,
    and illustrate that the whole DP phase  ($x_+ \!< \! x \!< \! \infty$) is critical and in the same universality for each $n$.} 
    \label{fig:ON_loop_phase}
\end{figure}

The phase diagram of the O($n$) loop model is shown in Fig.~\ref{fig:ON_loop_phase}.
As the bond weight $x$ is increased, 
the system undergoes a second-order phase transition $x_{+}$ from the dilute phase, 
consisting of  small loops, into a critical densely-packed (DP) phase,
where loops have fractal structures. 
The universality of the DP phase is governed by the line $x_{-}$ of stable fixed points. 
The exact locations of $x_\pm (n)$ are known as~\cite{nienhuis1982exact}
 \begin{equation}
 \label{eq:ON_solution}
    1/x_{\pm} = \sqrt{2 \pm \sqrt{2-n}} \; .
\end{equation}
The two curves meet at $(n \! = \!2,x_{\pm}\! =\! 1/\sqrt{2})$.
In the language of universality, the O($n$) loop model on the critical branch $x_{+}$ 
 (DP phase) corresponds to the tricritical (critical) $Q$-state Potts model 
 with $Q=n^2$ \cite{nienhuis1982exact,nienhuis1991locus}.
An interesting note is that, at $x_{+}$, the O(1) loop model is typically   
regarded as the critical Ising model instead of the tricritical one-state Potts model. 

In other words, despite of the absence of an exact transformation between the loop model (6) and the Potts model (1), the two systems have the same Coulomb-gas coupling strength $g$ as long as $Q=n^2$. This correspondence is expected to hold true for any real $0 \leq Q \leq 4$, where, for noninteger $Q$, the Potts model should be formulated in the FK bond representation.
The relation between $g$ and $Q$ is still given by Eq.~(\ref{eq:CG-coupling}), 
with $g \in [2,4]$ for $x_{-}$ and $g \in [4,6]$ for $x_{+}$. 
Furthermore, Eqs.~(\ref{eq:CGprediction1}) and (\ref{eq:CGpredictions}) 
for critical exponents still hold true.  
However, what physical quantities in the loop model, described by these equations, 
need to be identified, 
and this has been largely explored in Refs.~\cite{deng2007cluster,liu2011worm}.
The thermal field along the $x$ direction 
has exponent $y_{\rm t2}$ in Eq.~(\ref{eq:CGprediction1}), instead of $y_{\rm t1}$. 
The loops and the domains (faces) enclosed by the loops 
have the fractal dimensions $d_{\rm hull}$ and $\df=y_{\rm h1}$, respectively. 
The quantities, of which the scaling is governed by $y_{\rm t1}$, are identified 
as the magnetization and the susceptibility of Ising spins living on 
the triangular lattice that is dual to the honeycomb lattice. 

Highly efficient worm-type and cluster algorithms have been developed 
for the O($n$) loop model with $n \geq 1$~\cite{deng2007cluster,liu2011worm}.
It is shown that critical slowing-down barely exists at $x_{+}$, and is completely absent in the DP phase,
as qualitatively understood as following.  For worm-type and cluster algorithms, 
the dynamic exponent $z$ has the so-called Li-Sokal bound \cite{li1989rigorous} $z \geq z_{\rm min}=\alpha/\nu$, 
where $\alpha/\nu$ is the scaling exponent for specific heat, 
and, further, this lower bound is rather sharp $z \approx z_{\rm min}$ in 2D.   
For the critical Potts model, one has $z_{\rm min} \! = \! 2 y_{\rm t1}-2$, 
leading to $z \! > \! 0$ for $Q \! > \! 2$. 
Moreover, as $Q$ increases, the critical slowing-down becomes more severe, 
since $y_{\rm t1}$ is an increasing function of $Q$.
However, the O($n$) loop model has $\alpha/\nu=2 y_{\rm t2}-2$, 
and thus, $z_{\rm min} (n \geq 1) \leq 0$ along $x_{+}$   
and $z_{\rm min} \! < \! -2$ in the whole DP phase. 

We aim to determine the backbone exponent $\dB$ and the shortest-path exponent $\dm$ 
by studying the O($n$) loop model along the line $x_{-}(n)$. 
For this goal, we assume that the domains of DP loops and the corresponding FK random clusters 
not only have the same fractal dimension, but also exhibit the same scaling 
for other geometric properties. 
In comparison with the critical Potts model, the advantage of studying the loop model 
along $x_{-}(n)$ is twofold:
The absence of critical slowing-down and the absence of finite-size corrections from $y_{\rm t2}$  
(corrections from other sources can still exist). 

Our final estimates of $\dB$ and $\dm$ are given in Table~\ref{tab:estimate_result},
and significantly improve over the existing results. 
The agreement with the FK bond representation 
confirms our assumption that the domains of the DP O($n$) loops have the same geometric structures 
as the critical FK random clusters, 
and we expect that this correspondence can be extended to the $x_{+}$ branch.  

In the analysis of quantities associated with bridge-free (backbone) clusters 
for the DP O($n$) loop model, 
we observe that, despite the absence of corrections from $y_{\rm t2}$, 
finite-size corrections are still significant and become more and more severe as $n$ increases.
For the special case of $n=2$, logarithmic corrections seem to arise.
On the basis of our numerical results, we conjecture an exact formula, as a function of $n$,
for the leading correction exponent.

The remainder of this paper is organized as follows. 
Algorithms for simulation and measurement are described in Sec.~\ref{ Algorithms and Sampled Quantities},
together with a list of sampled quantities. 
Section~\ref{Results_backbone} presents our results for the backbone exponent 
and Sec.~\ref{Results_shortest-path} presents results for the shortest-path exponent. 
A brief discussion is given in Sec.~\ref{Conclusion}. 

  \section{Algorithms and Observables}
  \label{ Algorithms and Sampled Quantities}

	\subsection{Simulation of the Potts model }

For integer $Q$, the celebrated SW 
and Wolff cluster algorithms are used,  
and, for real $Q \geq 1$, the Chayes-Machta (CM) method~\cite{chayes1998graphical} is applied. 
They can be understood by the so-called {\it induced-subgraph} picture as following Ref.~\cite{deng2007cluster}. 

For the RC model~(\ref{eq:RC}) on graph $G=(V,E)$, one first decomposes $Q$ as $Q = Q_\alpha \! + \! Q_\beta$.
Independently for each cluster in a FK bond configuration, 
one then chooses color ``$\alpha$'' with probability $Q_\alpha/Q$ or color ``$\beta$'' 
with probability $Q_\beta/Q$, 
and assigns the chosen color to all sites in the cluster. 
Consequently, the lattice sites are partitioned as $V \! = \! V_\alpha \cup V_\beta$,
and the edge set $E$ becomes $E=E_{\alpha} \cup E_{\beta} \cup E_{\alpha \beta}$;
an edge $e \in E_{\alpha\beta}$ connects a pair of sites, respectively, in $V_\alpha$ and $V_\beta$,
and it is not occupied by definition.
It can be seen that, conditioning on this decomposition, 
the bond configuration is nothing other than the combination of a $Q_\alpha$-state RC model
on the induced subgraph $G_\alpha =(V_\alpha, E_{\alpha})$ and another $Q_\beta$-state RC model on $G_\beta$.
Now, one can update these induced RC models by any valid algorithm. 
One valid update is the identity operation, which is ``do nothing'' for the RC model, 
corresponding to the ``inactive" color of Chayes and Machta~\cite{chayes1998graphical}. 
Of course, we must also include at least one nontrivial update.

A simple choice is to set $Q_\alpha=1$ for $Q \geq 1$,
so that the corresponding induced model on $G_\alpha$ 
is the standard bond percolation and can be trivially updated, i.e., 
each edge $e \in E_{\alpha}$ is independently occupied with probability $p$. 
On this basis, a basic variant of CM algorithm is formulated as: 
\begin{enumerate}
\item Independently for each cluster, choose the active color ``$\alpha$" with probability $1/Q$ 
      and otherwise ``$\beta$'' (inactive), resulting in a random partition $V=V_\alpha \cup V_\beta $
      as well as $E=E_{\alpha} \cup E_{\beta} \cup E_{\alpha \beta}$. 
\item Independently for each edge $e \in E_{\alpha}$, place an occupied bond with probability $p=v/(1+v)$ 
      according to Eq.~(\ref{eq:RC}), and for $e \in E \! \setminus \! E_{\alpha}$, 
      do nothing. 
\end{enumerate} 
Here $E \setminus E_\alpha$ represents a complementary subset of $E_\alpha$ with edges in $E_\alpha$ being excluded. A new FK bond configuration is then obtained after 
ignoring/discarding the randomly assigned colors. 

For $Q \geq 2$, the efficiency of the CM algorithm can be further improved by 
increasing the number of active colors to be $m=[Q]$, with $[Q]$ the ground integer of $Q$.
Namely, for each cluster, an active color $\alpha_i \; (i=1,\cdots,m)$ is chosen 
with probability $1/Q$, leading to $m$ copies of the bond percolation on $G_{\alpha_i}$.
For integer $Q$, where the ``{\it do-nothing}" operation is absent, 
the celebrated SW algorithm is recovered. 
A single-cluster variant of the CM method is also available \cite{deng2009single}.

\subsection{Simulation of the O($n$) model} 
The induced subgraph picture can also provide a versatile platform to 
design Monte Carlo algorithms for the O($n$) model~(\ref{eq:ON_loop_PF}) with $n \geq 1$. 
The basic idea is to set $n=n_\alpha+n_\beta$ with $n_\alpha=1$, 
and to simulate the induced O(1) loop model. 
More precisely, independently for each loop, one randomly assigns 
the active color ``$\alpha$''  with probability $1/n$ 
and the inactive color ``$\beta$" with probability $1-1/n$.
In addition, one assigns the active color to all those sites that are not on loops,
since they have an implicit weight of ``1" (unity), as in Eq.~(\ref{eq:ON_loop_PF}).
The O(1) loop model on the induced subgraph can then be simulated by the worm algorithm. 
A variant of worm-type algorithm for the O$(n)$ loop model in 2D and 3D can be found in Refs.~\cite{liu2011worm,liu2012n}.

\begin{figure}[t]
    \centering  
    \includegraphics[width=0.50\textwidth]{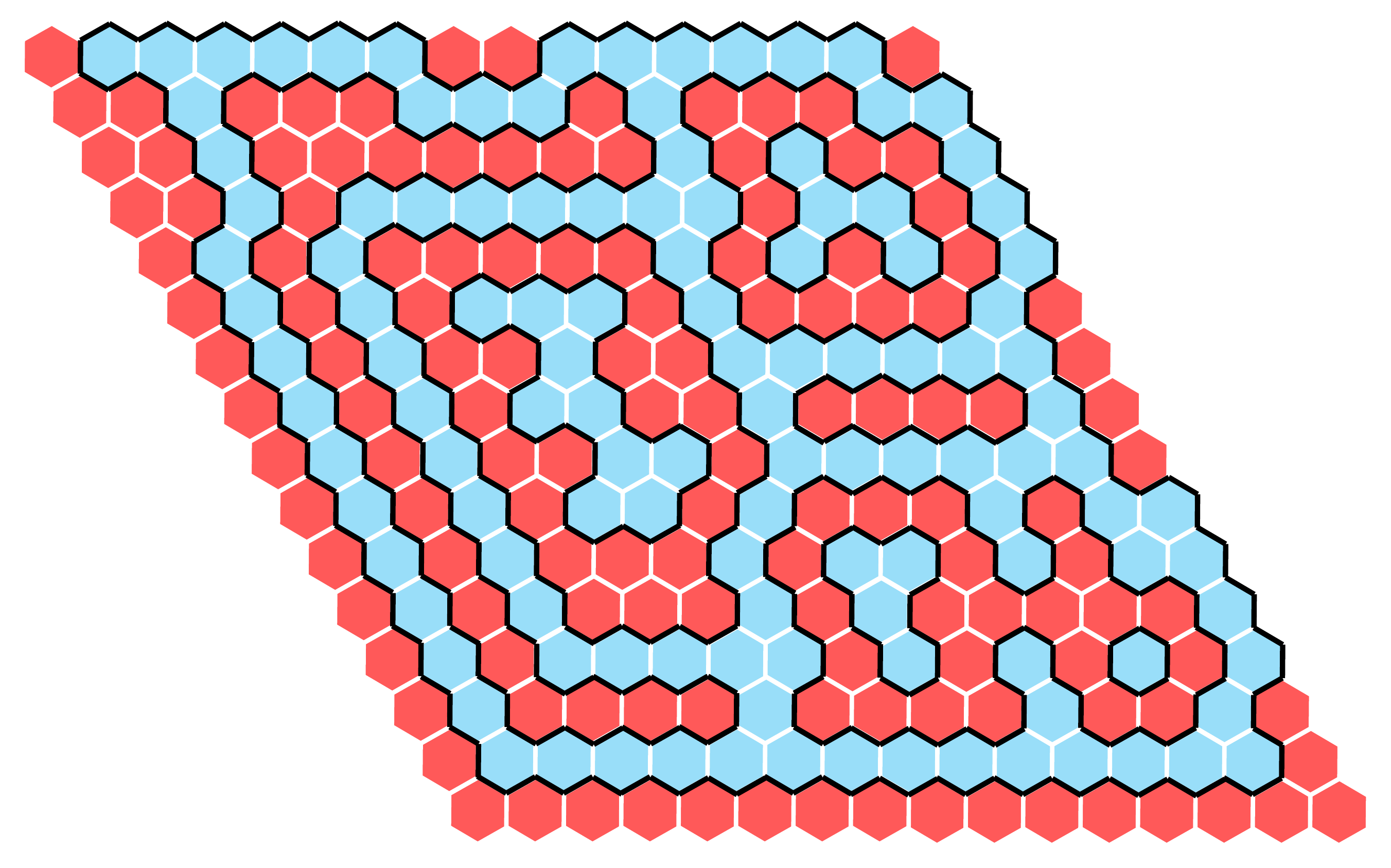}
    \caption{Illustration of the generalized Ising model on the triangular lattice and the O($n$) loop model on the honeycomb lattice with periodic boundary conditions. The hexagons in red (blue) denote the spin-up (spin-down) sites and loops, i.e., the domain walls between the generalized Ising spins,  are specified by  blue lines.}
    \label{fig:ONGIsn}
\end{figure}

Note that the honeycomb-lattice loops are the boundaries of Ising domains on the dual triangular lattice,
with the one-to-two correspondence between the loop and the spin configuration of dual Ising spins.
The dual coupling strength $K^*$ relates to $x$ as $2K^* \! = \! - \ln (x)$,
which is ferromagnetic for $x \! < \! 1$ and antiferromagnetic for $x \! > \! 1$.
The number of dual Ising domains is simply $\scrN_d = \scrN_\ell+1$, with $\scrN_\ell$ the loop number
and, thus, $n$ is also the fugacity of an Ising domain. 
In other words, the O$(n)$ loop model can be regarded as a generalized Ising model on the triangular lattice,
of which the partition function is written as 
\begin{equation}
\scrZ_{\rm GIsn} = \sum_{\{s \}}  n^{\scrN_d} \prod_{\langle i j \rangle } \exp(K^* s_i s_j ) \; ,
\label{eq:PF-GIsn}
\end{equation}
where the summation is over all the Ising configurations.
 Figure~\ref{fig:ONGIsn} demonstrates the relation between the loop model and generalized Ising model.
 The Ising spinslive on the sites of the triangular lattice,and the spin values are specified by the colors of the hexagonal faces. The domain walls are drawn between the boundaries of domains, which corresponds to the loops in the O$(n)$ loop model on the honeycomb lattice. In Fig.~\ref{fig:ONGIsn}, there are  $\scrN_d=11$ domains and $\scrN_\ell=10$ loops. 


Applying the induced-subgraph picture to Eq.~(\ref{eq:PF-GIsn}), one obtains a standard Ising model on 
induced subgraph $G_\alpha$ of the triangular lattice, which can be simulated by the SW or Wolff cluster 
algorithms. The bond-occupation probability is $p =  1\!-\! e^{-2K^*} \! = \! 1-x$ 
between a neighboring pair of parallel spins for $x \! \leq \! 1$ 
or $p=1\!-\!e^{2K^*} \! =\!1\!-\!1/x$ between anti-parallel spins for $x \!> \!1$.   
Nevertheless, some cautions are needed for the identify operation 
to the complementary subgraph $G \! \setminus \! G_\alpha$, 
which can no longer be ``{\it do-nothing}''.
Instead, one should place occupied bonds on all the edges $e \in E \! \setminus \! E_{\alpha}$, 
independent of the underlying Ising spins,
so that all sites in $V_\beta$ and their neighboring sites in $V_\alpha$ 
are in the same cluster. 
As a result, the domain topology on $G \! \setminus \! G_\alpha$
is kept unchanged, thanks to the Ising symmetry. 
For explicitness, the bond occupation probability for $x \! \leq \! 1$ is listed as 
\begin{equation}
	p = 
	\left \{
	\begin{array}{lcl}
		1-x  &     & \text{if $s_i = s_j$ and $i \in V_\alpha, j \in V_\alpha$;} \\
		0    &     & \text{if $s_i \neq s_j$ and $i \in V_\alpha, j \in V_\alpha $;} \\
		1    &     & \text{otherwise.}\\
	\end{array}
	\right.
\end{equation}  
For $x\! > \! 1$, similar procedure can be applied and the only difference is as follows: Within the active subgraph $G_\alpha$, the bond is placed between antiparallel spins ($s_i \neq s_j$) with probability $p=1-1/x$.

For the O$(n)$ loop model in the DP phase, both the cluster algorithm 
and the worm method exhibit no critical slowing-down, 
and their efficiency is approximately the same.
Since we are interested in the geometric structures of Ising domains in this work, 
simulating the generalized Ising model~(\ref{eq:PF-GIsn}) can avoid the complication 
of passing back and forth between the honeycomb and the triangular lattice.

\subsection{Identification of nonbridges}
Following Ref.~\cite{xu2014geometric},
we classify the occupied bonds of a FK bond configuration into bridges and nonbridges,
and study the resulting bridge-free clusters to determine the backbone dimension $\dB$. 
A bridge is an occupied bond whose deletion would break a cluster into two,
and removing all the bridges would produce a bridge-free configuration, 
which consists of isolated sites and blobs. 
In other terminology, blobs are also called bridge-free clusters and biconnected clusters, 
in which any two sites are connected by at least two independent paths.
For the Ising domains on the triangular lattice, 
occupied bonds have to be first placed within the domains before the bond classification.

Two distinct methods to identify nonbridges are 
introduced in Refs.~\cite{xu2014geometric} and~\cite{huang2018critical}, 
of which the former makes use of the planarity of 2D lattices. 
In the latter, clusters are grown by breadth-first procedure with help of a treelike data structure;
whenever a newly visited bond is to close a loop, 
one backtracks the two branches of tree until the joint site,
and all passing bonds are marked as nonbridges. 
A problem is that a nonbridge may be visited for many times. 
Since the number of nonbridges is of system size order, 
which is observed in Refs.~\cite{xu2014geometric,hou2019geometric}, 
a higher efficient algorithm is desired to identify nonbridges.

\begin{figure}[t]
    \centering  
    \includegraphics[width=0.50\textwidth]{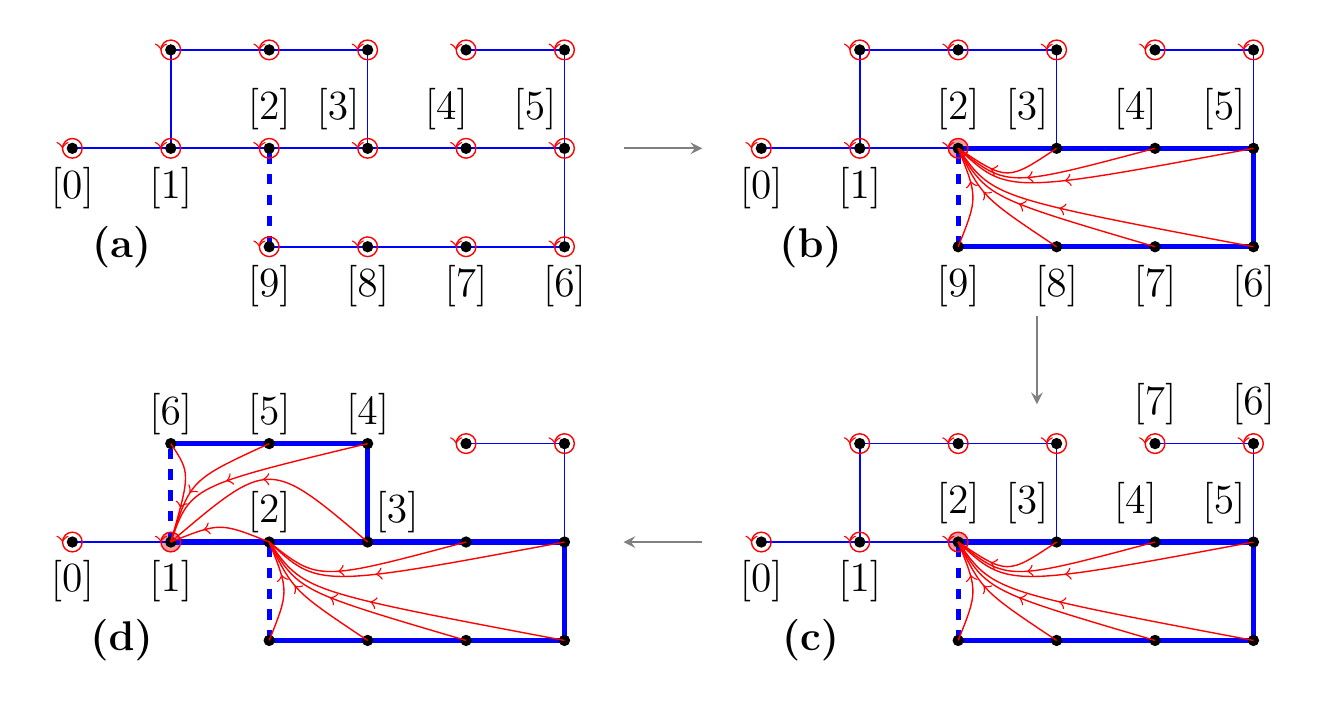}
    \caption{Identification procedure for nonbridges. 
    (a), Grow the depth-first oriented walk untill it is a to-be-closed loop, 
         as labeled by the walk $[0],[1], [2],\cdots,[9]$. 
    (b), Backtrack to the root site ($[2]$), construct a flat tree (red arrows), 
         and mark all the passing bonds as nonbridges (thick blue bonds). 
    (c), Apply the last-in-first-out rule so that  
         the old elements $[9],\cdots,[6]$ are removed and 
         and new elements, $[6]$ and $[7]$, are added to the walk. 
    (d), Add a new loop and replace the existing root by the new one. 
    The final configuration consists of four bridge-free clusters, 
    three of which are isolated sites (those with red circles). 
    }
    \label{fig:birdge-free}
\end{figure}

Hereby, we employ a depth-first growth with a chainlike data structure $\scrW [\ell ]$
and introduce a treelike graph $\scrD[i]$ with some minimal depth.
The chainlike array $\scrW [\ell ]$ is to keep trace of the depth-first growth:
Let $\scrW [\ell ]$ denote the newly added site, $\scrW[\ell\! + \! 1]$ stores the next to-be-added site.
It is required that, at one step, only {\it one} of the unvisited sites connecting to $\scrW[\ell]$ is added,
and thus the chain, $\scrW[0]$, $\scrW[1], \cdots, \scrW[\ell \! + \! 1]$, is nothing other than an oriented walk.  
Further, the ``last-in-first-out" rule is applied: If all neighboring sites of $\scrW[\ell]$ 
have been visited, then $\scrW[\ell]$ is removed and 
the incoming site $\scrW[\ell \! - \! 1]$ becomes the walk frontier.  
Array $\scrD[i]$ is a directed graph consisting of many trees, 
and each tree is to represent a bridge-free cluster (including an isolated site). 
The parent site of site $i$ is given by $\scrD[i]$, 
and the root of a tree is the site with the parent being itself. 
Initially, $\scrD[i]=i$ is set for all sites $i \in V$.

The procedure is sketched in Fig.~\ref{fig:birdge-free}. 
Starting with an initial site, the depth-first walk is performed 
until it is a to-be-closed loop, labeled by $[0], [1], \cdots, [9]$ in Fig.~\ref{fig:birdge-free}(a).
Then, the root is found starting from the joint site, [2],
and the loop is identified by backtracking $\scrD[i]$ (if applicable) or $\scrW[\ell]$ to the root.
Meanwhile, all the passing bonds are marked as nonbridges (thick blue bonds) 
and all the passing sites take the root as their parent [Fig.~\ref{fig:birdge-free}(b)]. 
Afterwards, the depth-first walk is continued with the last-in-first-out rule, 
as illustrated by Fig.~\ref{fig:birdge-free}(c). 
When a new loop is added, the existing root is taken place by the new one 
if they are different [Fig.~\ref{fig:birdge-free}(d)]. 
In this way, it is kept that the parent $\scrD[i]$ of site $i$ 
is always visited no later than site $i$ itself, 
so that the loop identification by backtracking $\scrD[i]$ works properly.
When all bonds in the cluster are visited, the depth-first walk is done,
and all the nonbridges are successfully identified. 
As a by-product, all bridge-free clusters are constructed
so that if one can further keep track of the sizes of trees, 
the associated quantities can be calculated immediately.    

In the backtracking procedure for identifying a loop, 
$\scrD(i)$ has a higher priority than  $\scrW[\ell]$: 
If $\scrD(i) \neq i$, then the backtracking is applied to $\scrD[i]$ instead of $\scrW[\ell \! - \! 1]$.
This is to avoid visiting a nonbridge for many times. 
In this work, we observe that trees $\scrD(i)$ are very flat
and their typical depth is of a few units. 
In practice, the trees can be further flatten: 
When a loop is to be formed, one tries to continue the walk by 
visiting another neighboring site of $\scrW[\ell]$. 
In other words, a self-avoiding walk is performed as long as possible 
before the backtracking procedure. 
Therefore, the present method has a nearly optimal efficiency for identifying nonbridges.

{\bf Sampled quantities.}--On the basis of bridge-free clusters, 
we sample the size of the largest cluster as  $B_1 = \langle \scrB_\textsc{1} \rangle$ 
and calculate the second moment of cluster sizes as $B_2 = L^{-2} \langle \sum \scrB^2 \rangle $,
where the sum is over all bridge-free clusters $\scrB$ and $\langle \cdot \rangle $
represents the ensemble average.
For the O$(n)$ loop model, we also distinguish 
the spin-up domains from the spin-down ones, 
and calculate the averaged size $B_\textsc{a} = \langle (\scrB_{1\uparrow}+\scrB_{1\downarrow})/2 \rangle$. 

To determine the shortest-path exponent $\dm$, we use the breadth-first algorithm to grow clusters;
the number $\scrS$ of layers after finishing the cluster growth 
corresponds to the maximum graph distance between the seed site and any other site in the cluster. 
We then sample the longest graph distance among all the clusters as $S_1 = \langle \scrS_1 \rangle$. 

At criticality, these sampled quantities are expected to asymptotically diverge
as a power-law scaling of linear system size $L$. More precisely, they behave as 
\begin{eqnarray}
S_1 &\sim& L^{\dm} \;, \hspace{15mm}  \mbox{and} \nonumber \\
B_1 & \sim &  B_\textsc{a} \sim  L^{\dB} \; , \hspace{8mm}  B_2 \sim L^{2\dB -d} \; .
\label{eq:exponent}
\end{eqnarray}

\section{Results for backbone exponent}
\label{Results_backbone}
We simulate the critical $Q$-state Potts model on the square lattice 
and the DP O$(n)$ loop model with system size $L$ ranging from 4 to 8192. 
Periodic boundary conditions are implemented.
For the Potts model, we take $Q=2,3$ and the exact critical point $v_c=\sqrt{Q}$. 
For the loop model, the simulation is at the branch $x_{-}(n)$ 
with $n=1, \sqrt{2} ,\sqrt{3}, \sqrt{2+\sqrt{3}}, 2$,
corresponding to $Q=1, 2, 3, 2+\sqrt{3}, 4$, respectively. 
 We generate more than $10^8$ independent samples for each system size $L < 256$,
and no fewer than $10^7$ independent samples for $L\ge 256$. \par

We perform least-squares fits to the finite-size scaling ansatz
\begin{equation}
\label{eq:fitting_ansatz}
    \scrO = L^{y_{\scrO}}(a + b_1 L^{y_1} + b_2 L^{y_2}) + c_0  \; ,
\end{equation}
where $\scrO= S_1, B_1, B_\textsc{a}, B_2$ 
and the corresponding exponent $y_{\scrO}$ is given by Eq.~(\ref{eq:exponent}). 
The term with $c_0$ describes the background contribution of the system, 
and those with $b_1$ and $b_2$ account for leading and subleading finite-size corrections, respectively. 
In comparison with the leading scaling term with amplitude $a$, 
all the other three terms play a role of corrections.  
As a precaution against correction-to-scaling terms that we miss including in the fitting ansatz, 
we impose a lower cutoff $L \ge L_{\rm m}$ on the data points admitted in the fits, 
and systematically study the effect on the residuals (denoted by $\chi^2$) by increasing $L_{\rm m}$. 
In general, the preferred fit for any given ansatz corresponds to the smallest $L_{\rm m}$ 
for which the goodness of the fit is reasonable and for which subsequent increases in $L_{\rm m}$ do not cause the $\chi^2$ value 
to drop by vastly more than one unit per degree of freedom (DF).
In practice, by “reasonable” we mean that $\chi^2/{\rm DF}  \approx  1$. 
The systematic error is obtained by comparing estimates from various reasonable fitting ansatz. \par 

\begin{table}[t]
    \centering
    \begin{tabular}{|c|lllllr|}
    \hline 
    $Q$ & $L_{\rm m}$  	&$\dB$ 	&$a$ 	&$b_1$ 	&$y_1$ 	& $\chi^2/{\rm DF}$ 	\\
    \hline 
    2   & 8     &1.732\,17(9)	&0.7995(6) 	&0.1168(9) 	&-0.66(1)		&4.0/7\\ 
        & 16    &1.732\,0(1) 	&0.8003(8) 	&0.121(4)  	&-0.68(2)  		&2.1/6\\ 
        & 32    &1.732\,0(2) 	&0.800(1)  	&0.12(1)   	&-0.68(5)  		&2.1/5\\ 
        &16     &1.732\,23(8)	&0.7990(5) 	&0.104(3)  	&-5/8	     	&2.3/6\\ 
        &32     &1.732\,1(1) 	&0.7995(8) 	&0.098(7)  	&-5/8		    &1.5/5\\ 
        \hline 
    3   & 16    &1.794\,0(6) 	&0.638(7)  	&0.415(2)  	&-0.44(1)  		&10.0/6\\ 
        & 32    &1.793\,6(9) 	&0.64(1)   	&0.420(9)  	&-0.46(3)  		&9.8/5\\ 
        & 16    &1.793\,3(1) 	&0.647(1)  	&0.415(2)  	&-0.46     		&12.0/7\\ 
        & 32    &1.793\,5(2) 	&0.645(2)  	&0.421(4)  	&-0.46     		&9.8/6\\ 
        & 64    &1.793\,2(3) 	&0.648(3)  	&0.411(10) 	&-0.46     		&8.6/5\\ 
        \hline 
        \end{tabular} 
        \caption{Fitting results of $B_1$ for the Ising model 
                 and of $B_2$ for the $Q=3$ Potts model.}
\label{tab:Potts_dB} 
\end{table}

\subsection{Estimate from the Potts model}
The $Q\! \rightarrow \! 1$ Potts model is the standard bond percolation, 
and the geometric structures of critical percolation clusters on the square lattice 
have been extensively studied in Ref.~\cite{xu2014geometric}, 
which gives $\dB=1.643 \, 36(10)$. 

\begin{figure}[b]
\centering
\includegraphics[scale=1.15]{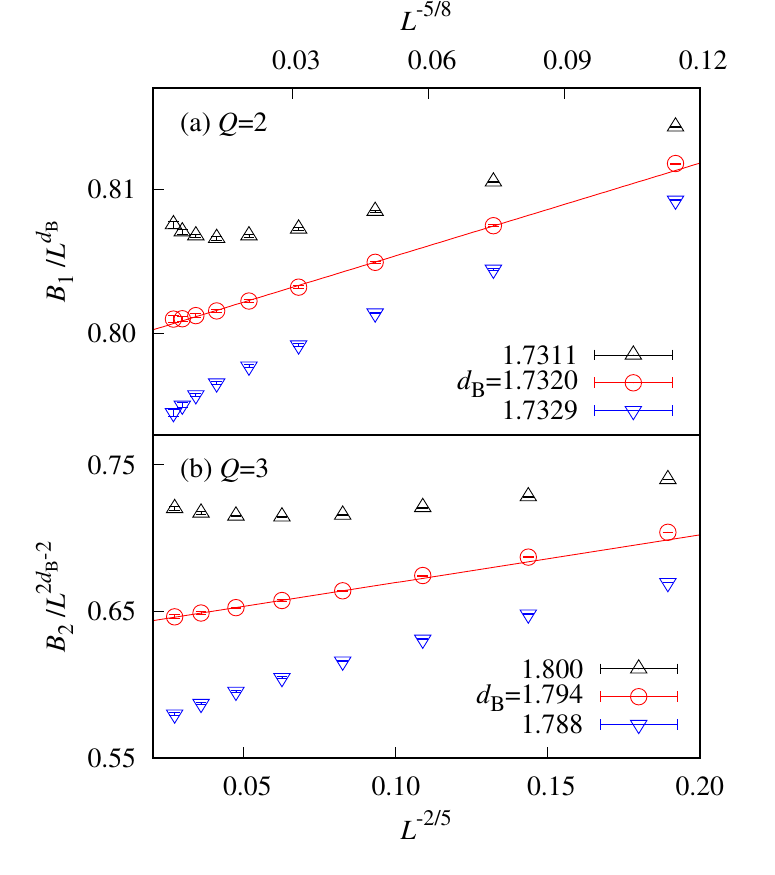}
\caption{Estimated backbone exponent $\dB$ for the Ising model (a) and the $Q=3$ Potts model (b).
 The approximate linearity of red lines indicates that 
 the leading correction exponent is $y_1 \approx -0.6$ for $Q=2$ and $-0.4$ for $Q=3$.
 The upward and downward bending of other curves  
 reflects the reliability of the finally quoted central values and their error bars. 
 }

\label{fig:Pottsq23}
\end{figure}

\begin{figure}[b]
\centering
\includegraphics[scale=1.15]{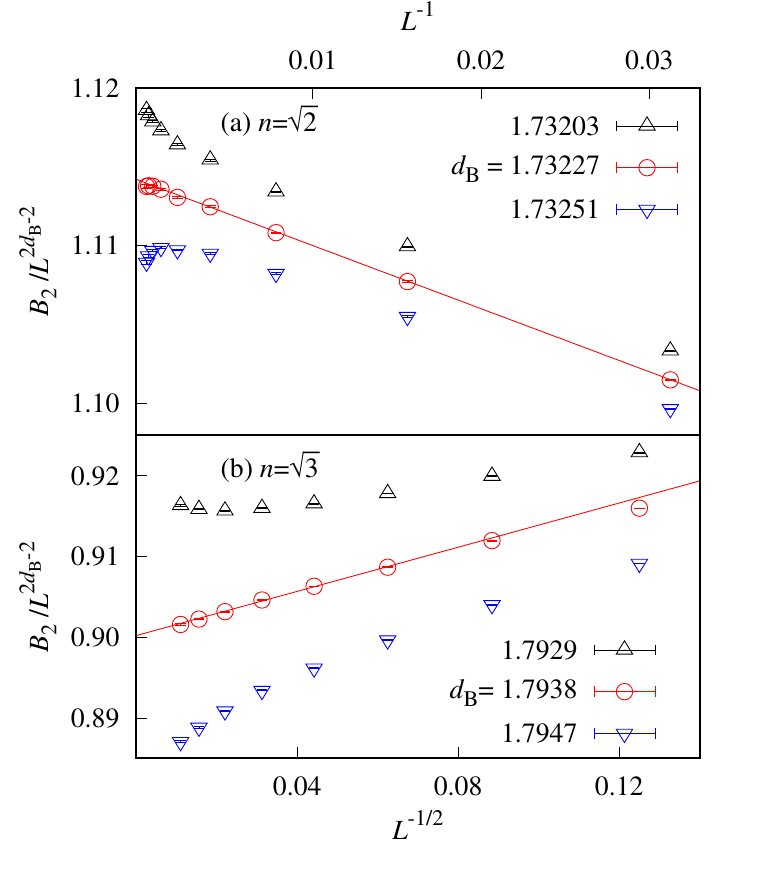}
\caption{Backbone exponent $\dB$ for the DP O($n$) loop model 
  with $n=\sqrt{2}$ (a) and $n=\sqrt{3}$ (b). 
  The upward or downward bending of curves with the $\dB$ value deviating from 
  the estimated value illustrates the reliability of the quoted error bars. 
  By comparing to Fig.~\ref{fig:Pottsq23},  
  it is seen that the results from the loop model have higher precision.}
\label{fig:ONdBq2}
\end{figure}

The Coulomb-gas coupling of the Ising model ($Q=2$) is $g=3$, 
and, according to Eq.~(\ref{eq:CGprediction1}),
the subleading thermal and magnetic scaling fields have exponents $y_{\rm t2} =-4/3$
and $y_{\rm h2}=13/24$, respectively. 
However, the subleading magnetic field with $y_{\rm h2}$ 
is considered to be redundant due the $Z_2$ symmetry of the Ising spins, 
and the subleading thermal field with $y_{\rm t2}$ is believed to play no role 
in the scaling of thermodynamic quantities in the spin representation.  
Indeed, finite-size corrections with exponent $y_{\rm t2} =-4/3$ 
are not observed in physical observables;
for quantities like energy, magnetization, specific heat, and susceptibility, 
the leading finite-size corrections are found to be governed by exponent $-2$.
Nevertheless, it remains open whether corrections with exponent $-4/3$ 
would arise in some geometric quantities.

According to the least-squares criterion, the $B_1$ data for $Q=2$ are fitted to Eq.~(\ref{eq:fitting_ansatz}). 
At first, we set $b_2=0$ and leave $\dB$, $a$, $b_1$, $y_1$ and $c_0$ as free parameters, 
but no stable fitting results can be obtained. 
By further fixing $c_0 = 0$, which is effectively a correction term, 
we obtain $\dB = 1.732\,0(2)$ and $y_1 = -0.68(5)$, 
of which the correction exponent is much larger than $-2$ or $-4/3$. 
The results are shown in Table~\ref{tab:Potts_dB}. 
Therefore, for geometric quantities associated with bridge-free clusters, 
the leading finite-size corrections probably arise from some other resource. 
In the framework of conformal field theory, an exponent $-5/8$ appears 
at several places in the Kac table for the Ising model \cite{Cardy1987}. 
With $y_1=-5/8$ being fixed, the fit yields $\dB = 1.732\, 23(8)$ for $L_{\rm m}=16$ 
(Table~\ref{tab:Potts_dB}). 
Following the same procedure, the fit of the $B_2$ data gives $\dB = 1.732\,3(5)$, 
which is consistent with $\dB$ from $B_1$ but has a slightly larger error bar. 
Comparing the fitting results of $B_1$ and $B_2$ for different $L_{\rm m}$,
we take the final estimate as $\dB = 1.732\,0(3)$.
We choose not to seriously take into account the smaller error bar
from the fits with $y_1=-5/8$ being fixed,
since it is only a crude guess and the exact value of $y_1$ is not available. 
In Fig.~\ref{fig:Pottsq23}(a), the $B_1/L^{\dB}$ data are plotted versus $L^{-5/8}$, 
with $\dB$ chosen to be the central value of the estimate as well as the central value plus 
or minus three error bars. 
The approximate linearity of the red line and the upward and downward bending curves
reflect the reliability of the final estimate $\dB=1.732 \, 0$
and of the quoted error margin $0.000 \, 3$.

The leading correction exponent for the $Q=3$ Potts model, 
with the Coulomb-gas coupling $g=10/3$, is $y_{\rm t2}=-4/5$. 
By fixing $b_2=0$ for the subleading correction term and $c_0=0$ for the background contribution, 
we fit the $B_2$ data to Eq.~(\ref{eq:fitting_ansatz}) and obtain  $\dB = 1.793\,6(10)$ and $y_1 =-0.46 (3)$.
The value of $y_1$ is much larger than $y_{\rm t2}=-4/5$, again suggesting that 
the leading finite-size corrections arise from some other source. 
The data are also well described when the term with $c_0$ is included 
and $y_1=-0.46$ is fixed in Eq.~(\ref{eq:fitting_ansatz}), 
as shown in Table~\ref{tab:Potts_dB}. 
Corrections due to the subleading thermal field with $y_{\rm t2}=-4/5$ should also exist, 
and, thus, there simultaneously exist corrections from different sources. 
Nevertheless, to identify different correction terms
is very challenging for numerical analysis, particularly because our data are limited.
The fits of $B_1$ yield $\dB = 1.792(2)$.
To be conservative, we quote our final estimate as $\dB = 1.794(2)$, 
of which the reliability is demonstrated by Fig.~\ref{fig:Pottsq23}(b). 

We did not simulate the Potts model for larger value of $Q$, 
due to severe critical slowing-down and finite-size corrections.

\subsection{Estimate from the loop model}
\label{Estimate for ON loop model}
For the O($n$) loop model, the backbone dimension $\dB$ can be obtained from 
the finite-size scaling of quantities including $B_1, B_\textsc{a}$, and $S_2$,
all of which give results consistent with each other. 
Nevertheless, the estimates from $B_2$ has somewhat better precision than those $B_1$ 
and $B_\textsc{a}$, probably because $B_2$ includes the sizes of all Ising domains in its definition. 
Table~\ref{tab:ONq_S2} summarizes the fitting results from $B_2$ for $n=1, \sqrt{2},
\sqrt{3}, \sqrt{2+\sqrt{3}}$, corresponding to $Q=1, 2, 3, 2+\sqrt{3}$. 
The fitting details are given below.

\begin{table}[b]
\centering
\scalebox{1.0}{
\begin{tabular}{|c|lllllr|}
\hline 
$Q$ & $L_{\rm m}$  	&$\dB$ 	&$a$ 	&$b_1$   	&$y_1$  & $\chi^2/{\rm DF}$ 	\\
\hline 
 1&   16    &1.643\,39(1)	&1.4922(2) 	&-1.529(9)    &-1.082(3)        &4.2/6\\ 
     & 32    &1.643\,40(2)	&1.4918(4) 	&-1.57(4)     &-1.091(9)        &3.0/5\\ 
     &64    &1.643\,39(2)	&1.4921(6) 	&-1.5(1)      &-1.08(2)         &2.8/4\\ 
\cline{2-7}
     &32    &1.643\,35(1)	&1.4932(3) 	&-1.07(1)     &-1    		&4.2/5\\ 
     &64    &1.643\,37(2)	&1.4928(4) 	&-1.02(3)     &-1     	        &2.1/4\\ 
\hline 
   2 &32    &1.732\,27(3)	&1.1140(5) 	&-0.40(3)     &-1.00(3)  	&5.4/4\\ 
     &64    &1.732\,27(5)	&1.1139(8) 	&-0.4(1)      &-1.01(8)  	&5.3/3\\ 
\cline{2-7} 
     &32    &1.732\,27(1)	&1.1140(2) 	&-0.399(3)    &-1        	&5.4/5\\ 
     &64    &1.732\,27(2)	&1.1140(3) 	&-0.400(8)    &-1        	&5.4/4\\ 
\hline 
 3   &16    &1.793\,9(1) 	&0.898(3)  	&0.105(4)     &-0.42(3)  	&5.4/5\\ 
     &32    &1.793\,8(2) 	&0.900(4)  	&0.11(1)      &-0.46(6)  	&4.9/4\\ 
\cline{2-7} 
     &32    &1.793\,66(3)	&0.9024(5) 	&0.126(3)     &-1/2      	&5.4/5\\ 
     &64    &1.793\,70(5)	&0.9018(8) 	&0.130(5)     &-1/2      	&4.4/4\\ 
\hline 
  2  &8     &1.838\,3(2) 	&0.757(3)  	&0.210(2      &-0.304(6) 	&4.5/6\\ 
$+$  &16    &1.838\,4(2) 	&0.755(5)  	&0.211(2)     &-0.30(1)  	&4.3/5\\ 
\cline{2-7}
$\sqrt{3}$  
     &8     &1.838\,38(3)	&0.7546(4) 	&0.2114(7)    &-0.3     	&5.0/7\\ 
     &16    &1.838\,36(4)	&0.7549(5) 	&0.211(1)     &-0.3     	&4.3/6\\ 
     &32    &1.838\,32(5)	&0.7555(8) 	&0.209(2)     &-0.3     	&3.5/5\\ 
\hline
\end{tabular} }
\caption{Fitting results of $\dB$ from $B_{2}$ for the DP O$(n)$ loop model 
     with $n =1, \sqrt{2} ,\sqrt{3}, \sqrt{2+\sqrt{3}}$.}
\label{tab:ONq_S2} 
\end{table}

\subsubsection{$n \! = \! 1$} 
From Eq.~(\ref{eq:ON_solution}) for the line of stable fixed points, 
one has the bond weight $x\! =\! 1$ for $n\! =\! 1$, meaning that the coupling strength for the dual Ising 
model $K^*\! =\! -\frac{1}{2}\ln x \! = \! 0$
and thus the Ising spins on different lattice sites 
are independent. With up (down) spins being interpreted as occupied (empty) sites, 
the DP O(1) loop model at the branch $x_{-}$ is just the site percolation on the triangular lattice. 
This is also reflected in the induced-subgraph cluster algorithm, as formulated in Sec.~II, 
where the bond occupation probability $p=1-x=0$ 
and each site is randomly occupied or empty with probability $50\%$.

\begin{figure}[t]
\centering
\includegraphics[scale=1.15]{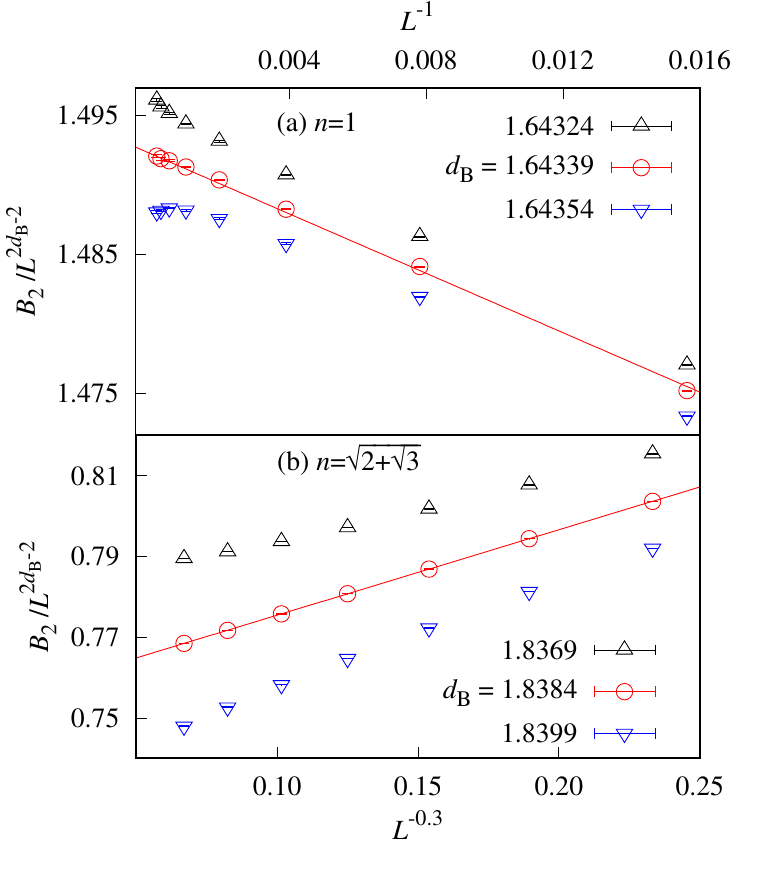}
\caption{Estimated backbone exponent $\dB$ from the domains of the DP O($n$) loop model 
with $n=1$ (a) and $n=\sqrt{2+\sqrt{3}}$ (b). 
The O(1) loop model, which has the bond weight $x=1$, is actually the site percolation on the triangular lattice. 
The upward or downward bending of those curves, with the $\dB$ value deviating from 
the estimated value, illustrates the reliability of the quoted error bars. }
\label{fig:ONdBq1}
\end{figure}

In the least-squares fit of the second moment $B_2$ of bridge-free clusters by Eq.~(\ref{eq:fitting_ansatz}),
if $b_2=0$ and $b_1$, $y_1$ and $c_0$ are left as free parameters, 
then we have $\dB=1.643\,38(4)$ but the correction amplitude $b_1$ 
is consistent with zero within its error bar. 
With further fixing $c_0=0$, we obtain $\dB = 1.643\,39(2)$ and $y_1 = -1.09(1) $. 
The $B_2$ data can be also well described by Eq.~(\ref{eq:fitting_ansatz}) 
with $b_1=b_2=0$ and $c_0$ as free parameter, and the backbone dimension is $\dB=1.643\,35(5)$.
The fitting results are listed in Table~\ref{tab:ONq_S2}.
By considering all different fits, we take the final estimate as $\dB = 1.643\,39(5)$,
of which the reliability is illustrated in Fig.~\ref{fig:ONdBq1}. 
The fits for $B_1$ and $B_\textsc{a}$ give $\dB =1.643\,37(9)$ and $\dB = 1.643\,38(5)$, 
respectively, which are consistent with that from $B_2$ but have slightly larger error bars.

\subsubsection{$n \! = \! \sqrt{2}$}
As described in Sec.~II, 
the current study of the honeycomb-lattice O($n$) loop model 
is to simulate the generalized Ising model Eq.~(\ref{eq:PF-GIsn}) on the triangular lattice 
by the induced-subgraph cluster method. 
To determine the backbone dimension $\dB$ for $Q=2$,  
one can either simulate the standard Ising model at the critical branch $1/x_{+}=\sqrt{3}$
and study the FK random clusters constructed in the cluster simulation or simulate the O($\sqrt{2}$) loop model in the DP phase and study the Ising domains.
Namely, within the framework of the loop model, there exists a correspondence between 
the FK random clusters constructed at the critical branch $x_{+}$ with $n=1$
and the Ising domains in the DP phase with $n^*=\sqrt{2}$.

We argue that such a geometric correspondence can be extended to general $n$.
Given the honeycomb-lattice loop model with parameters $(n,x)$, one can always construct 
on the triangular lattice the Ising domains and FK-like random clusters 
by placing occupied bonds with probability $1-x$ within each domain. 
Along the critical branch $x_+(n)$, 
we assume the following duality relation: 
Both the fractal dimensions of domains and of FK-like random clusters, 
$d_\textsc{face}$ and $d_\textsc{fk}$, 
are given by $y_{\rm h1}$ in Eq.~(\ref{eq:CGprediction1});
nevertheless, the Coulomb-gas coupling $g$ for $d_\textsc{face}$ 
is in $[4,6]$ and calculated from $n$ as $n=-2 \cos (\pi g/4)$, 
while $g_*$ for $d_\textsc{fk}$ is in $[2,4]$ and relates to $g$ as $g_*=16/g$.
As a consequence of this duality relation, 
we further argue that the FK-like random clusters for $(n,x_+)$ 
have the same geometric structures as the domains of the DP O($n_*$) loop model 
with $n_* =-2 \cos (\pi g_*/4)$, 
similarly to the correspondence between the critical and tricritical $Q$-state  
Potts model observed in Ref.~\cite{deng2004geometric}. 
For the marginal case $n=n_*=2$, this means that, 
apart from multiplicative logarithmic corrections, 
the FK-like random clusters and the domains have the same fractal dimension.
Finally, it is noted that, for $n <2$, the FK random clusters in the DP phase 
are too small to percolate. 
A simple example is for $(n \! = \! 1,x_{-} \! = \! 1)$, where the bond probability is $p \! = \! 1-x=0$ 
and each FK-like cluster is just a single site.

For the critical Ising model, the induced-subgraph cluster method reduces to 
the SW cluster algorithm, which still suffers from some critical slowing-down 
due to the logarithmic divergence of the specific heat. 
Hereby, we determine the backbone dimension $\dB$ from the O$(\sqrt{2})$ loop model 
at the branch $x_-$, for which critical slowing-down is completely absent. 

We perform the least-squares fits for the $B_2$ data to Eq.~(\ref{eq:fitting_ansatz}).
By setting $b_2 \! = \! 0$ and leaving the other correction parameters free, 
we obtain $\dB  \! =  \! 1.732\,19(5)$, and $c_0$ is consistent with zero within its error bar. 
By further fixing  $c_0  \! = \! 0$, we have $\dB  \! =  \! 1.732\,27(5)$ and $y_1  \! =  \! -1.01(8)$.
The error bar of $\dB$ can be further suppressed if $y_1\!=\!-1$ is taken.
Nevertheless, this smaller error bar cannot be taken into account seriously,
since we are not aware from what sources such finite-size corrections arise.
The fitting results are given in Table~\ref{tab:ONq_S2}, and
the final estimate is quoted as  $\dB = 1.732\,27(8)$, 
of which the reliability for the central value and the quoted error bar is 
illustrated in Fig.~\ref{fig:ONdBq2}. 
The fits for $B_1$ and $B_\textsc{a}$ give $\dB =1.732\,2(1) $ and $\dB = 1.732\,0(3)$, respectively,
in excellent agreement with that from $B_2$.

\subsubsection{$n \! = \! \sqrt{3}$}
The same fitting procedure is applied to the data of $B_1, B_\textsc{a}, B_2$.
It is now necessary to simultaneously include two correction terms (e.g., with $b_1$ and $c_0$).
For $n \! = \! \sqrt{3}$, we obtain  $\dB = 1.793\,8(3)$ 
from $B_2$, $1.793\,5(7)$ from $B_1$ and $1.794\,0(7)$ from $B_\textsc{a}$.
The leading correction exponent is estimated as $y_1 = -0.42 (3) $, 
which is in good agreement with 
$y_1=-0.46 (3)$ from the $Q=3$ Potts model. 
Both estimates of $y_1$ are significantly larger than $y_{\rm t2}=-2/5$. 
This strongly suggests that the leading corrections associated with backbone clusters
are from some scaling field with exponent near $ \approx -0.4$. 
The fitting results are listed in Table~\ref{tab:ONq_S2},
the final estimate is taken as $\dB = 1.793\,8(3)$ and shown Fig.~\ref{fig:ONdBq2}(b).

\subsubsection{$n \! = \! \sqrt{2+\sqrt{3}}$}
The Coulomb-gas coupling is $g=11/3$,
leading to $y_{\rm t2} \! = \! -4/11  \! \approx \!  -0.364 $. 
With $b_2=0$, we obtain exponents $\dB = 1.838\,2 (2 )$ and $y_1 = -0.304 (6)$ for $L_{\rm m}=8$, 
and the correction amplitudes are also well determined as $b_1=0.210(2)$ and $c_0=-0.254(8)$. 
Although the estimated value of $y_1$ is larger than $y_{\rm t2}$ 
if the error bar of $y_1$ is taken into account, 
it is difficult to conclude $y_1 \neq y_{\rm t2}$.
Nevertheless, it is clear that, as $n$ increases, 
finite-size corrections become larger and larger, 
albeit they are smaller than those for the corresponding Potts model. 
As $n \rightarrow 2$, we expect that correction exponent $y_1 \rightarrow 0$ and 
multiplicative logarithmic corrections would arise.  
The final estimate is taken as $\dB = 1.838\,4 (5)$ and shown in Fig.~\ref{fig:ONdBq1}(b).

\subsection{Logarithmic corrections for $n=2$}
The two branches of the O($n$) loop model meet 
at $x_{\pm}=1/\sqrt{2}$ for $n=2$, 
and the thermal field associated with the bond weight $x$ becomes marginal (i.e., $y_{\rm t2}=0$).
In accordance with the phase diagram in Fig.~\ref{fig:ON_loop_phase}, 
this thermal field is marginally irrelevant for $x > x_{\pm}$ 
and marginally relevant for $x < x_{\pm}$, and, at $x_{\pm}$, its amplitude vanishes. 
It is expected that logarithmic corrections, irrespective of multiplicative or additive forms, 
are absent in thermodynamic quantities, including energy density, 
specific heat, and magnetic susceptibility,
as well as in geometric quantities associated with the sizes of loops or domains.

\begin{table}[t]
    \centering
    \begin{tabular}{|c|llllr|}
    \hline 
    $L_{\rm m}$  	&$y_{D_2}$ 	&$a$ 	&$b_1$ 	&$y_1$ 	& $\chi^2/{\rm DF}$ 	\\
    \hline 
    4     &1.750\,000(8)	&0.89188(4)	&0.079(4)  	&-2.00(4)  	&4.0/8\\ 
    8     &1.749\,993(10)	&0.89192(5)	&0.12(5)   	&-2.2(2)   	&2.7/7\\ 
    \hline
    4     &1.750\,000(6)	&0.89188(3)	&0.0787(6) 	&-2        	&4.0/9\\ 
    8     &1.750\,001(7)	&0.89187(3)	&0.079(2)  	&-2        	&4.0/8\\ 
    \hline 
    \end{tabular} 
    \caption{Fitting results of the second moment $D_2$ of domain sizes 
             for the O$(2)$ loop model at $x_{\pm}=1/\sqrt{2}$. }
    \label{tab:ON_q4_D2} 
    \end{table}
    
\begin{figure}[b]
\centering
\includegraphics[scale=0.71]{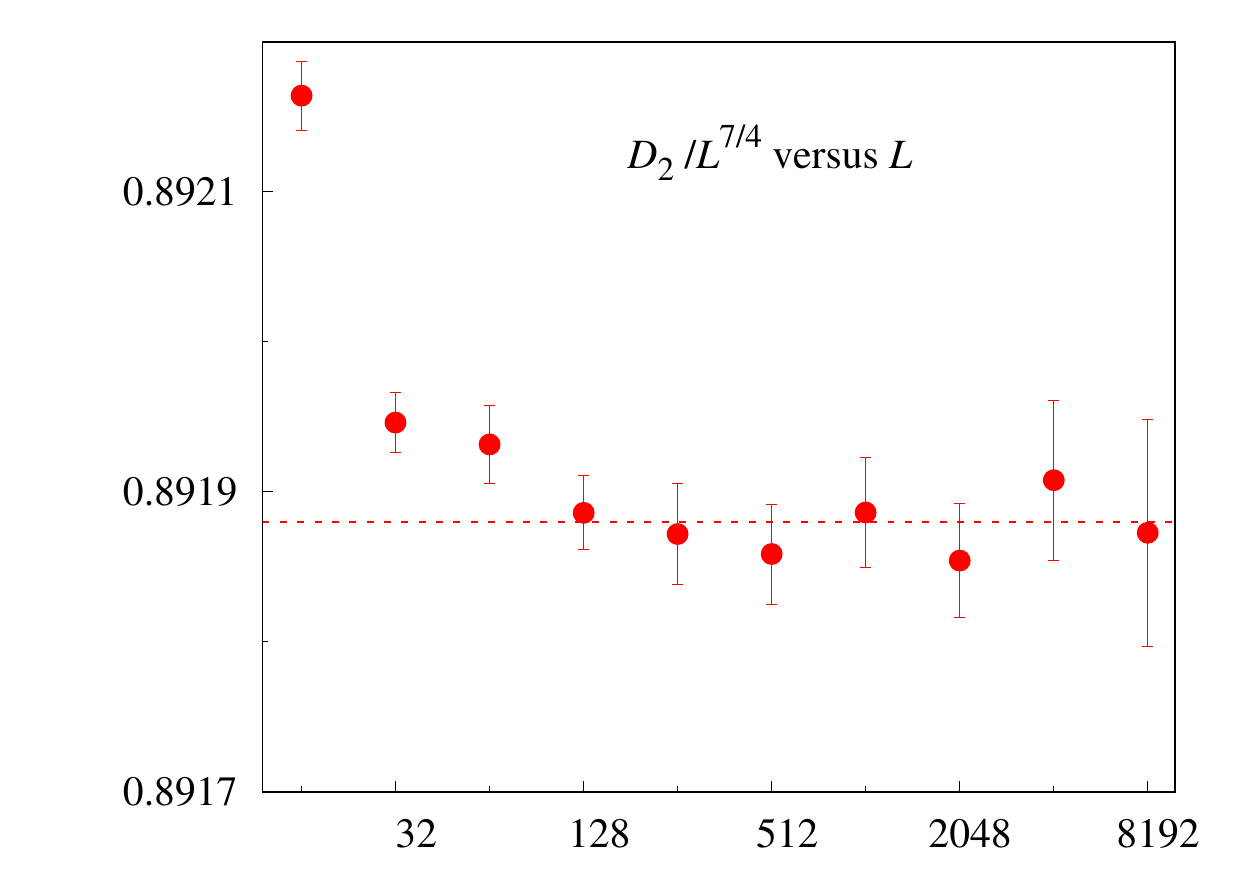}
\caption{Absence of multiplicative and additive logarithmic corrections 
  in geometric quantities associated with domain sizes 
  of the O$(2)$ loop model at $x_{\pm}=1/\sqrt{2}$, 
  as illustrated by the rescaled quantity $D_2/L^{7/4}$. 
  The horizontal axis is the log plot of linear size $L$. 
  All the finite-size data of $D_2/L^{7/4}$ are more or less 
  consistent with the horizontal red dashed line at $0.89188$, 
  strongly indicating that corrections of logarithmic types are absent 
  and those of algebraic forms are very small.
   }
\label{fig:ON_q4_D2}
\end{figure}

To confirm this expectation, we sample the second moment of 
domain sizes as $D_2 = L^{-2} \langle \sum \scrD^2 \rangle$,
where the summation is over all the dual Ising domains $\scrD$. 
The domains have a fractal dimension $\df=y_{\rm h1}$ given by Eq.~(\ref{eq:CGprediction1}), 
and, for $n=2$, it is expected $D_2 \sim L^{2\df-2} = L^{7/4}$ with $\df=15/8$.
By fixing $b_2=c_0=0$ so that 
only one single correction term is included in Eq.~(\ref{eq:fitting_ansatz}), 
we observe that the reasonably good fits can already be obtained with the minimum size 
as small as $L_{\rm m}=4$. 
The estimated exponent, $\df=1.875\,00(1)$, is excellent agreement with the exact value $15/8$.
The correction exponent $y_1=-2.00(4)$ is well consistent with $-2$,
and, furthermore, the correction amplitude $b_1 = 0.079(4)$ is very small;
see Table~\ref{tab:ON_q4_D2} for details. 
For clarity, the rescaled data of $D_2/L^{7/4}$ are plotted versus size $L$ 
in Fig.~\ref{fig:ON_q4_D2},
and it can be seen that, as long as for $L \geq 128$, 
finite-size corrections are buried in statistical noise. 
The rapid convergence of $D_2/L^{7/4}$ demonstrates the absence of multiplicative or additive 
logarithmic corrections for $n=2$. 
It also strongly supports that, along the line $x_{-}(n)$ of stable fixed points, 
finite-size corrections with $y_{\rm t2}$ should not occur 
in quantities associated with loops and domains. 

\begin{table}[t]
\centering
\scalebox{1.0}{
\begin{tabular}{|l|lllllr|}
\hline 
$\scrO$ &$L_{\rm m}$  &$\dB$ 	&$a$ 	&$b_1$    &$y_1$ 	& $\chi^2/{\rm DF}$ 	\\
\hline 
$B_2$   &16    &1.867\,1(3) 	&0.733(6)  	&0.230(3)  	&-0.29(1)  	&3.7/5\\ 
        &32    &1.866\,7(4) 	&0.740(9)  	&0.230(2)       &-0.30(2)  	&2.8/4\\ 
        &64    &1.866\,9(8) 	&0.74(2)   	&0.229(3)       &-0.29(4)  	&2.8/3\\
\hline
$B_\textsc{a}$   
        &32    &1.869(1)  	&0.522(9)  	&0.051(8)       &-0.20(4)  	&1.5/4\\ 
        &64    &1.868(1)       &0.531(10) 	&0.046(5)  	&-0.25(8)  	&0.9/3\\ 
\hline
$B_1$   &16    &1.866\,4(6) 	&0.780(6)  	&0.149(3)     	&-0.30(2)  	&3.7/5\\ 
        &32    &1.866\,2(9) 	&0.782(10) 	&0.149(2)       &-0.31(4)  	&3.6/4\\ 
\hline 
\end{tabular} }
\caption{Fitting results for backbone exponent $\dB$ for the O$(2)$ loop model without 
     logarithmic corrections--i.e., Eq.~\eqref{eq:fitting_ansatz}.}
\label{tab:ONq4} 
\end{table}

For the backbone exponent $\dB$, we first try to fit the data 
to Eq.~\eqref{eq:fitting_ansatz}, without logarithmic corrections being included,
and the results are shown Table~\ref{tab:ONq4}.
Taking $B_2$ for an example, if we fix $b_2=0$ and leave other parameters free, 
then no stable results can be obtained. 
When $c_0=0$ is further fixed, we have $\dB=1.866\,9(8)$. 
Similar analysis for $B_\textsc{a}$ and $B_1$ gives $\dB=1.868(2)$ and $1.866(2)$, respectively. 
As shown in Table~\ref{tab:ONq4}, all the estimates are consistent with each other. 
The estimated value of $\dB$ is slightly smaller than the fractal dimension $\df=15/8$.

In the earlier study \cite{deng2004backbone}, it was numerically observed that 
the backbone exponent $\dB$ of
the tricritical Potts model reduces to the fractal dimension $\df$ of FK random clusters,
which can be explained by the fact that the red-bond scaling field is 
irrelevant (i.e., $d_{\rm red}<0$ for $g>4$).
For the O(2) loop model with $g=4$, the red-bond scaling field is marginal $d_{\rm red}=0$,
and thus, $\dB=\df=15/8$ might still be expected. 
Nevertheless, 
the fitting results of $\dB$ in Table~\ref{tab:ONq4} deviate from the predicted value
systematically when the error bars are taken into account. 
A plausible reason is due to logarithmic corrections, 
which may still occur in backbone clusters. 
This is also implicitly reflected by 
the small correction exponent $y_1 \! \approx \! -0.2$, as shown in Table~\ref{tab:ONq4},
which might actually correspond to logarithmic corrections. 

\begin{table}[t]
\centering          
\setlength\tabcolsep{1.0pt}
\begin{tabular}{|l|llllllc|}           
\hline 
$\scrO$ &$L_{\rm m}$  	&$\dB$ 	&$a$ 	&$b$ 	&$\hat{y}_{\scrO}$ 	&$d_0$ 	 & $\chi^2/{\rm DF}$ 	\\
\hline 
$B_2$ &64    &1.875    	&1.171(3)  	&0.06(3)   	  	&-0.246(2) 	&1.928(4)        &2.3/3\\ 
      &128   &1.875    	&1.171(8)  	&0.07(9)   	   	&-0.246(4) 	&1.929(9)        &0.5/2\\ 
\cline{2-8} 
      &128   &1.875\,6(3)    &1.163(2)   &-0.12(8)    &-0.25 		&1.66(2)         &2.1/2\\     
\hline 
$B_{\rm A} $&16    &1.875    	&0.694(4)  	&-0.042(2) 	   	&-0.126(2) 	&4.2(1)    	  	&5.1/6\\ 
            &32    &1.875    	&0.700(7)  	&-0.038(5) 	   	&-0.129(3) 	&4.4(3)    	   	&4.3/5\\ 
\cline{2-8}
&16    &1.874\,93(8)	&0.6919(9) 	&-0.042(2) 	   	&-0.125   	&4.13(7)   	   	&5.0/6\\ 
&32    &1.874\,8(1) 	&0.693(2)  	&-0.039(4) 	   	&-0.125   	&4.2(1)    	   	&4.2/5\\ 
\hline
$B_1$ &16    &1.875    	&0.960(6)  	&-0.01(1)  	   	&-0.120(2) 	&0.7(1)    	   	&4.3/6\\ 
      &32    &1.875    	&0.956(10) 	&-0.03(3)  	   	&-0.118(3) 	&0.6(2)    	   	&4.0/5\\
      &64    &1.875    	&0.96(2)   	&-0.01(6)  	   	&-0.120(6) 	&0.7(4)    	   	&3.9/4\\ 
\cline{2-8} 
&16    &1.875\,3(1) 	&0.971(2)  	&-0.006(8) 		&-0.125   	&0.88(6)   	   	&4.3/6\\ 
&32    &1.875\,4(2) 	&0.970(3)  	&-0.01(2)  	  	&-0.125   	&0.8(1)    	   	&4.1/5\\ 
&64    &1.875\,3(3) 	&0.972(5)  	&0.01(4)   	 	&-0.125   	&0.9(2)    	   	&3.9/4\\ 
\hline 
\end{tabular} 
\caption{Fitting results of $\dB$ for the O$(2)$ loop model with 
   the logarithmic ansatz~\eqref{eq:fitting_ansatz_log} and $y_1=-1$. } 
\label{tab:ONq4_log} 
\end{table}

We assume that quantities on the basis of backbone clusters 
follow the following finite-szie scaling ansatz  
        \begin{equation}
 \label{eq:fitting_ansatz_log}
\scrO = L^{y_{\scrO}} [\ln L+ d_0]^{\hat{y}_{\scrO}} (a  + b L^{y_1} )  \; ,
\end{equation}
where the constant $d_0$ is commonly used in the fitting 
with multiplicative logarithmic corrections. 
In comparison with the general scaling ansatz for logarithmic corrections,
we have ignored in Eq.~(\ref{eq:fitting_ansatz_log})
additive logarithmic corrections which appear in powers of $1/\ln L$. 
This is because, in the fit, (i) the effects of parameter $d_0$ and of those additive 
logarithmic corrections can interfere with each other, 
and (ii) our data are limited and can already be reasonably described 
without multiplicative logarithmic corrections.

\begin{figure}[t]
\centering
\includegraphics[scale=0.7]{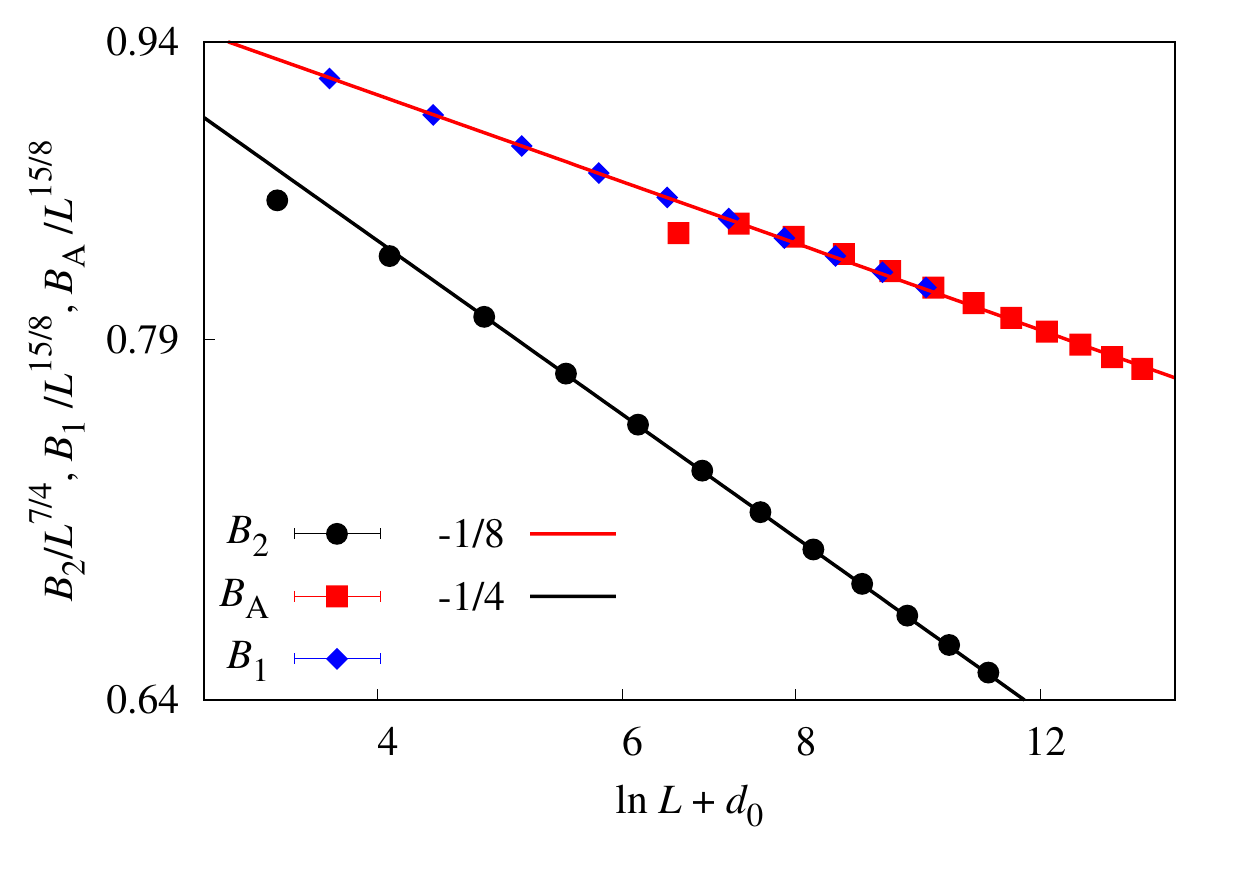} 
\caption{Log-log plot of rescaled backbone quantities versus $(\ln L + d_0)$ 
   for the O$(2)$ loop model. The rescaled quantities are $B_2/L^{7/4}$,
   $B_1/L^{15/8}$, and $B_\textsc{a}/L^{15/8}$. For being concise, $B_\textsc{a}$ is multiplied by 
   a constant such that the data for $B_\textsc{a}$ and $B_1$ approximately collapse on top of each other. 
   The existence of multiplicative logarithmic corrections 
   and the corresponding exponents are clearly illustrated. 
   }
\label{fig:ONq4}
\end{figure}

Since no prediction exists for the 
logarithmic exponent $\hat{y}_{\scrO}$ for backbone clusters 
and it is difficult to get an estimate to $y_{\scrO}$ 
and $\hat{y}_{\scrO}$ simultaneously, 
we first fix $y_{\scrO}$ with expected value and get the estimate of $\hat{y}_{\scrO}$.
For $B_2$, we first fix $y_{\scrO} = 2\dB - 2=7/4$ and leave all other parameters free,  which leads to no stable results. Then we fix $y_1=-1$ and obtain $\hat{y}_{\scrO} = -0.246(2)$ for $L_{\rm m}=64$.
More fits have been done with $y_1$ fixed to various values in the interval $[-2, -1)$, which give the estimate $\hat{y}_{\scrO}=0.24(1)$. 
Later, we fix $y_1 = -1$ and $\hat{y}_{\scrO}$ to various values in the range $-0.23\le \hat{y}_{\scrO} \le -0.25$, and we obtain the estimate $ \dB = 1.875\,3(6)$ by covering all the fitting results.
Following the same procedure, we obtain the estimate $\dB =1.874\,7(7)$ and $\hat{y}_{\scrO}= 0.135(15)$ for $B_{\rm A}$, $\dB=1.875\,2(7)$ and $\hat{y}_{\scrO} = 0.122(8)$ for $B_1$. The detail of the fitting results are summarized in Table.~\ref{tab:ONq4_log}. Thus, we expect $(y_{\scrO},\hat{y}_{\scrO})$ is $(15/8, 1/8)$ for $B_\textsc{a}$ and $B_1$, and is $(7/4, 1/4)$ for $B_2$.
This is further demonstrated in Fig.~\ref{fig:ONq4}, 
where the approximately straight lines in the log-log plot 
correspond to logarithmic exponents $\hat{y}_{\scrO}=-1/4$ and $-1/8$.

In the four-state Potts model,  
the multiplicative logarithmic correction exponent is $-1/16$ for 
the magnetization~\cite{salas1997logarithmic},
and thus, the size of the largest FK cluster
also has logarithmic exponent $-1/16$, 
half of our estimate $-1/8$ for backbone clusters. 
This is an interesting observation, 
but it might be due to the absence of additive logarithmic corrections 
in Eq.~\eqref{eq:fitting_ansatz_log}. 
Unfortunately, it is very challenging to obtain conclusive evidence 
from fits simultaneously with multiplicative and additive logarithmic corrections. 
Further theoretical insights are needed for the exact value of 
the logarithmic exponent for backbone clusters at $n=2$. 

\subsection{Conjectured formula for leading correction exponent in backbone clusters}
The fitting results of $\dB$ from the $Q$-state Potts and the DP O($n$) loop model 
suggest that, in geometric observables associated with backbone clusters,  
the leading correction exponent is $y_1 \! \approx \! -1.0, -0.6, 
-0.4, -0.2$ for $Q=1, 2, 3, 2+\sqrt{3}$, respectively [the estimate $y_1 \! \approx \!-1$ 
for the O($\sqrt{2}$) loop model might be due to the small amplitude of the correction term].
For $Q=4$ or $n=2$, there exists a marginal scaling field, 
leading to multiplicative and additive logarithmic corrections. 
These corrections cannot be accounted by the scaling field with $y_{\rm t2}$.
It is tempting to identify the exact value of $y_1$ 
from the Kac table in the conformal field theory \cite{henkel1999conformal}.

The Kac table is characterized by a pair of integers $i,j\geq 0$. 
The value of a scaling dimension $X_{i,j}$, 
which relates to the corresponding exponent $y$ as $y=2-X$, 
is expressed as \cite{friedan1984conformal,Cardy1987}
\begin{equation}
X_{i, j}= \frac{[i(m+1)-jm ]^2-1}{2m(m+1)} \; ,
\label{eq:CFT}
\end{equation}
where the conformal anomaly $m$ can be calculated from the Coulomb-gas 
coupling as $m=g/(4-g)$ for $g \in [2,4]$ and $m=4/(g-4)$ for $g \in [4,6]$. 
For the critical Potts model, this gives  $m=2, 3, 5, \infty$ 
for $Q \! = \! 1, 2, 3, 4$, respectively. 
The exactly known exponents, in Eqs.~(\ref{eq:CGprediction1}) and~(\ref{eq:CGpredictions})
can be mostly identified in the Kac table with integer $(i,j)$.
An exception is the leading magnetic dimension $X_{\rm h1}$, 
of which the indices $i\! =\! j \! = \! (m \! + \! 1)/2$ are half integers for $m=2$ (percolation).

On the basis of the estimated values of $y_1$, 
we conjecture that the corresponding indices are $(i,j)=(2,0)$, leading to 
\begin{eqnarray}
y_1 &=& -(4m+3)/2m (m+1) \; ,  \nonumber \\
    &=& -(4-g)(g+12)/8g \; .
\label{eq:CFT_X1}
\end{eqnarray}
This predicts $y_1 \!=\! -11/12, -5/8, -23/60, -47/263$ 
for $Q \! = \! 1, 2$, $3, 2 \! +\! \sqrt{3}$, respectively, 
which are surprisingly close to the numerical results of $y_1$. 
Moreover, Eq.~(\ref{eq:CFT_X1}) gives $y_1 \! \rightarrow \! 0$ 
as $g \! \rightarrow \! 4$, 
and thus can explain the appearance of logarithmic corrections for $n=2$.

\begin{table}[t]
\setlength\tabcolsep{1.5pt}
\centering
\begin{tabular}{|l|lllllc|}
\hline 
$Q$  &$L_{\rm m}$  	&$\dB$ 	&$a$ 	&$b_1$ 	&$b_2$ 		& $\chi^2/{\rm DF}$ 	\\
\hline 
$1$       &64    &1.643\,34(2)	&1.4935(5) 	&-0.67(3)  	&-5.8(8)   		&2.0/4\\ 
          &128   &1.643\,33(3)	&1.4937(8) 	&-0.69(7)  	&-5(4)     		&2.0/3\\ 
        \hline 
$2$       &16    &1.732\,20(2)	&1.1153(3) 	&-0.051(3) 	&-0.57(1)	  	&8.3/6\\ 
          &32    &1.732\,17(3)	&1.1159(6) 	&-0.059(6) 	&-0.53(3)  	  	&6.3/5\\ 
          &64    &1.732\,20(5)	&1.1154(9) 	&-0.05(1)  	&-0.6(1) 	  	&5.6/4\\ 
        \hline   
$3$       &64    &1.794\,1(1)	&0.894(2)  	&0.108(9)  	&-0.04(2)  	 	&6.8/4\\ 
          &128   &1.793\,8(2) 	&0.899(3)  	&0.07(2)   	&0.06(6)   	 	&3.2/3\\ 
          &256   &1.794\,0(3) 	&0.895(6)  	&0.10(4)   	&-0.0(1)   	 	&2.5/2\\ 
        \hline   
$2$       &32    &1.840\,0(3) 	&0.714(5)  	&0.14(1)   	&0.102(7)  	 	&1.6/5\\ 
$+$       &64    &1.840\,0(4) 	&0.715(9)  	&0.14(2)   	&0.10(1)   	 	&1.6/4\\ 
$\sqrt{3}$&128   &1.839\,3(7) 	&0.73(2)   	&0.10(4)   	&0.13(3)   	 	&0.3/3\\ 
\hline 
\end{tabular} 
\caption{Fitting results of $\dB$ from $B_2$ for the DP O($n$) loop model 
with $n=1,\sqrt{2},\sqrt{3},\sqrt{2+\sqrt{3}}$. 
The leading correction exponent $y_1$ is fixed at the value 
 given by Eq.~\eqref{eq:CFT_X1} and the subleading correction exponent is taken as $y_2=2y_1$.}
\label{tab:fit_ON_CFT} 
\end{table}

For the critical $Q$-state Potts model, the external-perimeter 
fractal dimension $d_\textsc{ep}$ has indices as $(i,j)=(1,0)$ in the Kac table, 
which is neighboring to $(i,j)=(2,0)$ conjectured for $y_1$. 
It is noted that $d_\textsc{ep}$ is also the fractal dimension 
of external perimeters and hulls of backbone clusters \cite{grossman1986structure}.
This observation indirectly supports that the conjectured formula~(\ref{eq:CFT_X1}) 
might be reasonable for backbone clusters.

On this basis, we reanalyze the $B_2$ data for the DP O$(n)$ loop model by Eq.~(\ref{eq:fitting_ansatz}) 
with $y_1$ being fixed at the predicted value.
Another correction term with $b_2$ or $c_0$ is also included,
and, for simplicity, we first fix the subleading correction exponent as $y_2=-2$. 
For $n=1$, the fit with $c_0=0$ gives  $\dB= 1.643\,33(3)$, 
and the fit with $b_2=0$ leads to $\dB = 1.643\,35(5)$, 
consistent with each other.
The same procedure has been done for $n=\sqrt{2}$ and $\sqrt{3}$,
and the results are consistent with the previous estimates.
For $n=\sqrt{2+\sqrt{3}}$, the residual $\chi^2$ is big until $L_{\rm m}=128$,
and the result is $\dB=1.841\,7(5)$ for $c_0=0$. 
When fixing $b_2=0$ and leaving $c_0$ free, we obtain $\dB = 1.841\,5(5)$. 
Both estimates deviate from the previous estimate $1.838\,5(4)$. 
This is probably due to that subleading finite-size corrections 
cannot be described by exponent $y_2=-2$.

We then set $y_2=2y_1$ and fix $c_0=0$. 
The final estimates $ \dB = 1.643\,33(3)$, $1.732\,20(6)$, $1.793\,9(4)$, 
$1.839\,5(9)$ are obtained  for $n = 1$, $\sqrt{2}$, 
$\sqrt{3}$, and $\sqrt{2 +\sqrt{3}}$, respectively, 
consistent with previous estimates. 
The fitting results with this choice are summarized 
in Table~\ref{tab:fit_ON_CFT}. 
For $n=\sqrt{2}$, we have $y_2=2y_1=-5/4$, close to $-1$. 
The amplitude  $b_1$ is much smaller than $b_2$, 
and thus, for small system sizes, the subleading corrections
can surpass the leading ones. 
This might explain that, in the previous fits, 
the estimated correction exponent is $y_1 \approx -1$ in Table~\ref{tab:ONq_S2}.
For small $n$, the precision of $\dB$ with this choice  
is slightly improved. 

\begin{table}[t]
\centering
\begin{tabular}{|c|lllllr|}
\hline 
$Q$  &$L_{\rm m}$  	&$\dm$ 	&$a$ 	&$b_1$ 	&$y_1$ 	 & $\chi^2/{\rm DF}$ 	\\
\hline 
              &32    &1.094\,42(7)	&1.1221(6) 	&-2.01(3)  	&-0.943(5) 	&2.3/4\\ 
              &64    &1.094\,5(1) 	&1.1212(9) 	&-2.10(8)  	&-0.96(1)  	&1.1/3\\ 
\cline{2-7}
$2$           &64    &1.094\,54(3)	&1.1209(3) 	&-2.132(7) 	&-0.96     	&1.2/4\\ 
              &128   &1.094\,53(5)	&1.1210(4) 	&-2.14(2)  	&-0.96     	&1.1/3\\ 
\hline 
              &32    &1.067\,45(9)	&1.0833(8) 	&-1.82(3)  	&-0.911(6) 	&2.7/4\\ 
              &64    &1.067\,5(1) 	&1.082(1)    	&-1.9(1) 	&-0.92(2)  	&2.1/3\\ 
\cline{2-7}
$3$           &32    &1.067\,58(3)	&1.0822(2) 	&-1.865(3) 	&-0.92     	&5.0/5\\ 
              &64    &1.067\,52(5)	&1.0826(4) 	&-1.878(9) 	&-0.92     	&2.1/4\\ 
              &128   &1.067\,52(7)	&1.0826(6) 	&-1.88(2)  	&-0.92     	&2.1/3\\ 
\hline 
              &32    &1.047\,4(1) 	&1.0332(9) 	&-1.71(3)  	&-0.907(8) 	&4.2/4\\ 
$2$	      &64    &1.047\,5(2) 	&1.033(2)   	&-1.7(1)   	&-0.91(2)  	&4.1/3\\ 
\cline{2-7}
$+$           &32    &1.047\,49(4)	&1.0328(3) 	&-1.724(4) 	&-0.91     	&4.4/5\\ 
$\sqrt{3}$    &64    &1.047\,47(6)	&1.0330(4) 	&-1.73(1)	&-0.91     	&4.1/4\\ 
              &128   &1.047\,49(9)	&1.0328(7) 	&-1.72(2)  	&-0.91   	&4.0/3\\ 
\hline       
              &32    &1.031\,9(1) 	&0.981(1)   	&-1.46(3)  	&-0.874(9) 	&4.1/4\\ 
$4$           &64    &1.032\,1(2) 	&0.980(2)   	&-1.6(1)   	&-0.90(2)  	&2.6/3\\ 
\cline{2-7}
	      &64    &1.032\,15(7)	&0.9795(5) 	&-1.59(1)	&-0.9      	&2.6/4\\ 
              &128   &1.032\,2(1) 	&0.9794(7) 	&-1.58(2)  	&-0.9      	&2.6/3\\ 
\hline 
\end{tabular} 
\caption{Fitting results of the shortest-path exponent $\dm$ from $S_1$ for the 
  DP O($n$) loop model with $n=\sqrt{2},\sqrt{3},\sqrt{2+\sqrt{3}},2$.}
\label{tab:fit_ON_dm} 
\end{table} 

\section{Results for shortest-path exponent}
\label{Results_shortest-path}

Since the shortest-path exponent $\dm$ of percolation ($Q \! \to \! 1$) 
is already estimated to have a high precision~\cite{zhou2012shortest}, 
we hereby present the results from the domains of the DP O($n$) loop model 
with $n \! = \! \sqrt{2}, \sqrt{3}, \sqrt{2 +\sqrt{3}}$, and 2.
We perform the least-squares fits to the $S_1$ data by the ansatz~\eqref{eq:fitting_ansatz}.
For $n=\sqrt{2}$, we first leave all parameters free but cannot 
get stable results. Then, we fix $c_0=b_2=0$ and obtain the estimate 
$\dm=1.094\,48(13)$ and $y_1 \approx -0.96$, close to $-1$. 
If we fix $y_2=-2$ and leave $b_2$ free, then the fit gives the consistent 
estimate as $\dm=1.094\,5(1)$ and $y_1=-0.97(2)$. 
In addition, fixing $y_1=-1$ and leaving $c_0$ free gives 
the estimate $\dm = 1.0945\,5(1)$. 
We take the final estimate $\dm = 1.094\,5(2)$ by considering 
the systematic error from different fits. 
Following the same procedure for $n=\sqrt{3}, \sqrt{2 +\sqrt{3}}, 2$,  
we obtain $\dm = 1.067\,5(3), 1.047\,5(3), 1.032\,2(4)$, respectively. 
The details of fits are summarized in Table~\ref{tab:fit_ON_dm}.

\begin{figure}[t]
\centering
\includegraphics[scale=1.10]{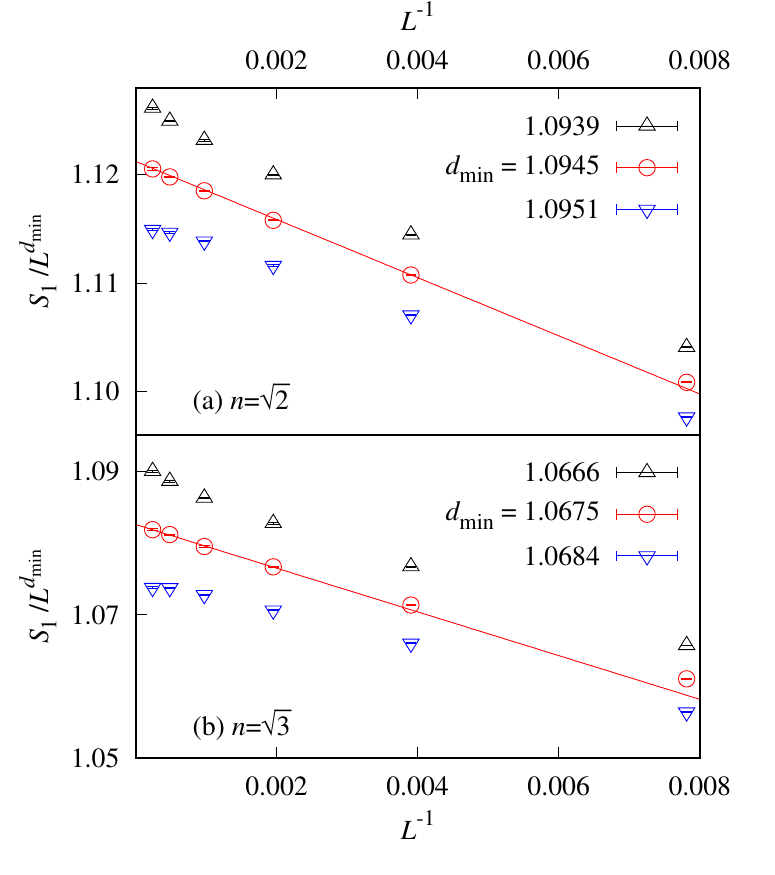}
\caption{Estimated shortest-path exponent $\dm$ from 
  the domains of the DP O($n$) loop model with $n=\sqrt{2}$ (a) and $n=\sqrt{3}$ (b). 
  The upward or downward bending of those curves, 
  with the $\dm$ value deviating from the estimated value, 
  illustrates the reliability of the quoted error bars.}
\label{fig:ON_dm_n23}
\end{figure} 

\begin{figure}[t]
    \centering
    \includegraphics[scale=1.10]{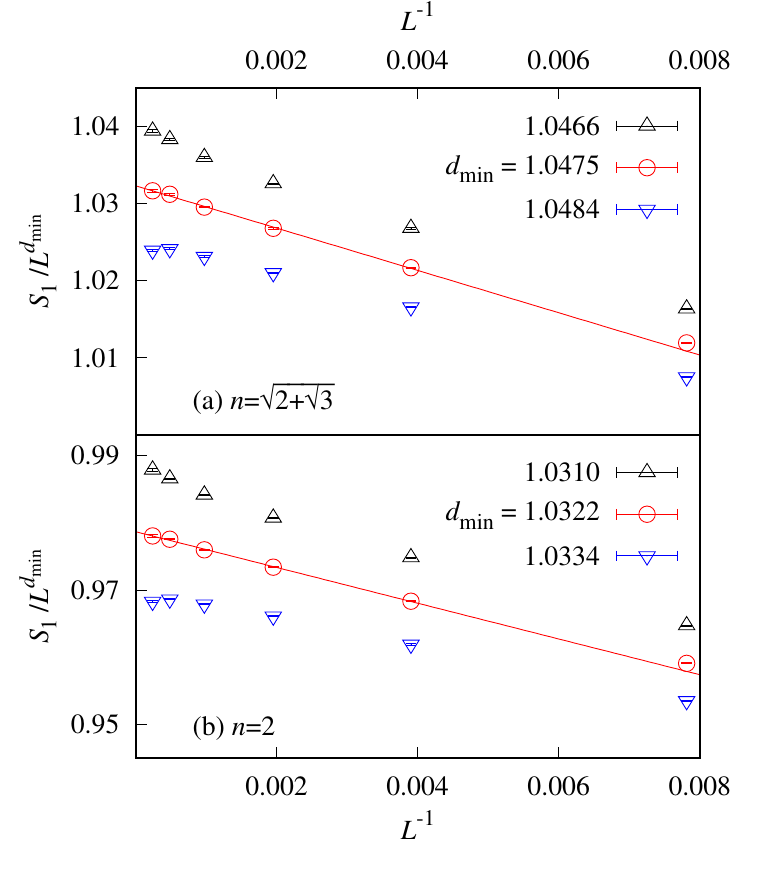}
    \caption{Estimated shortest-path exponent $\dm$ from the domains 
    of the DP O($n$) loop model with $n=\sqrt{2 + \sqrt{3}}$ (a) and $n=2$ (b). 
    The upward or downward bending of those curves, with the $\dm$ value 
    deviating from the estimated value, illustrates the reliability of the quoted error bars.}
    \label{fig:ON_dm_nR4}
    \end{figure}

As shown in Table~\ref{tab:fit_ON_dm}, 
we find that the leading correction exponents $y_1$ is always close to $-1$, 
irrespective of the value of $n$, and, thus, that
the precision of $\dm$ remains approximately unchanged as $n$ increases.  
Unlike for the backbone exponent $\dB$, finite-size corrections 
do not become stronger as $n$ increases.
In particular, no logarithmic corrections seem to arise for $n=2$. 
In Figs.~\ref{fig:ON_dm_n23} and \ref{fig:ON_dm_nR4}, 
we plot $S_1/L^{\dm}$ versus $L^{-1}$. 
The approximate linearity of the red line and the upward 
and downward bending curves reflect the reliability of 
the estimates and the quoted error bars.

\section{Discussion}
\label{Conclusion}
By cluster Monte Carlo methods, we carry out extensive simulations 
of the $Q$-state Potts model on the square lattice and 
of the O$(n)$ loop model on the honeycomb lattice, up to linear size $L=8192$.
The induced-subgraph picture adds a valuable perspective to understand 
the celebrated Swendsen-Wang algorithm and further provides a versatile platform to 
formulate efficient cluster or worm-type algorithms for loop models.  

The backbone exponent $\dB$ and the shortest-path exponent $\dm$ 
are determined for the Fortuin-Kasteleyn random clusters of the critical Potts model 
and for the domains of the DP O($n$) loop model.
According to Coulomb gas theory, the DP O$(n)$ loop model corresponds to 
the critical $Q$-state Potts model with $Q=n^2$.
The excellent agreement of the $\dB$ and $\dm$ results between the two systems 
strongly supports that, in addition to the overall fractal structure, 
the domains of the DP loop model and 
the corresponding FK random clusters share the same scaling for other geometric properties.

In comparison with the $Q$-state Potts model, the study of the DP O$(n)$ loop model 
benefits from the advantages that the absence of critical slowing down and 
the absence of finite-size corrections with exponent $y_{\rm t2}$ in Eq.~(\ref{eq:CGprediction1}).
Also, an improved algorithm for identifying backbones 
is formulated, by adopting a depth-first cluster growth procedure 
and introducing an auxiliary array of treelike structure.

As a consequence of these advantages, the estimates of $\dB$ and $\dm$
achieve a precision better than the existing results. 
It can be seen that, as $Q$ increases, the backbone exponent $\dB$ increases 
and the shortest-path exponent $\dm$ decreases, and,
for $Q=4$, the backbone exponent and the fractal dimension 
coincide (apart from logarithmic corrections).  
This reflects that the critical FK random clusters become more and more compact as $Q$ increases. 
Since the exact values of $\dB$ and $\dm$ are still unknown for 
the two-dimensional Potts model, our high-precision results can provide a solid testing ground 
for future theoretical explorations.

Finite-size corrections in the DP O$(n)$ loop model 
are also studied along the line of stable fixed points.
It is confirmed that corrections with exponent $y_{\rm t2}$ are absent.
Particularly, for the marginal case $n=2$, 
multiplicative and additive logarithmic corrections are not observed in 
quantities associated with the domain sizes and in the shortest path.
The leading correction exponent $y_1$ for the shortest path is observed 
to be roughly around $-1$, independent of $n$.
For backbone clusters, however, 
finite-size corrections become larger and larger as $n$ increases,
and logarithmic corrections arise for $n=2$. 
To account for these corrections, which are beyond the description of exponent $y_{t2}$, 
we conjecture an exact formula, Eq.~(\ref{eq:CFT_X1}), for the correction exponent $y_1$. 
The predicted value of $y_1$ is consistent with our Monte Carlo data, 
and can explain the emergence of logarithmic corrections for $n=2$.
Fixing the $y_1$ value can lead to smaller error bars of $\dB$, 
which are, however, not taken in our final estimate.
It remains as an interesting question to explore whether our conjecture holds true.

\section{Acknowledgments}
Y.D. acknowledges support by the National Natural Science Foundation of China (under Grant No. 11625522),
the Science and Technology Committee of Shanghai (under grant No. 20DZ2210100), and the National Key R\&D
Program of China (under Grant No. 2018YFA0306501). W.Z. acknowledges support by the National Natural Science Foundation of China Youth Fund (under Grant No. 12105133) and  the Fujian Provincial Natural Science Foundation of China (under Grant No. 2021J011030).

\end{document}